\documentclass{aa}
\usepackage{graphicx} 
\usepackage{dcolumn}  
\usepackage{bm}       
\usepackage{amssymb,amsmath}
\usepackage{placeins}

\usepackage{color}
\usepackage{float}
\usepackage{natbib}
\def\cgfit{\textsc{gFIT}}
\def\gfit{\textsc{gFIT}}
\def\galsim{\textsc{Galsim}}
\def\pse{\textsc{SExtractor}}
\def\ksb{\textsc{KSB}}
\def\shapelens{\textsc{shapelens}}

\newcommand{\Fref}[1]{Fig. \ref{#1}}

\begin{document}

\title{Shear measurement bias I: Dependencies on methods, simulation parameters, and measured parameters}
\titlerunning{Shear measurement bias}

\authorrunning
{Arnau Pujol\and
Florent Sureau\and
Jerome Bobin\and
et al.
}
\author
{Arnau Pujol \inst{1,2} \and
Florent Sureau \inst{1,2} \and
Jerome Bobin \inst{1,2} \and
Frederic Courbin \inst{3} \and \\
Marc Gentile \inst{3} \and
Martin Kilbinger \inst{1,2,4} \\
}
\institute{
DEDIP/DAP, IRFU, CEA, Universit\'e Paris-Saclay, F-91191 Gif-sur-Yvette, France
\and
AIM, CEA, CNRS, Université Paris-Saclay, Université Paris Diderot, Sorbonne Paris Cité, F-91191 Gif-sur-Yvette, France
\and
Institute of Physics, Laboratory of Astrophysics, Ecole Polytechnique F\'ed\'erale de Lausanne (EPFL), Observatoire de Sauverny, 1290 Versoix, Switzerland\\
\and
Institut d'Astrophysique de Paris, UMR7095 CNRS, Universit\'e Pierre \& Marie Curie, 98 bis boulevard Arago, 75014 Paris, France\\
}

\date{Received date / Accepted date}

\abstract{
We present a study of the dependencies of shear bias on simulation (input) and measured (output) parameters, noise, point-spread function anisotropy, pixel size, and the model bias coming from two different and independent galaxy shape estimators. We used  simulated images from \galsim\ based on the GREAT3 control-space-constant branch, and we measured shear bias from a model-fitting method (\cgfit) and a moment-based method (Kaiser-Squires-Broadhurst). We show the bias dependencies found on input and output parameters for both methods, and we identify the main dependencies and causes. Most of the results are consistent between the two estimators, an interesting result given the differences of the methods. We also find important dependences on orientation and morphology properties such as flux, size, and ellipticity. We show that noise and pixelization play an important role in the bias dependencies on the output properties and galaxy orientation. We show some examples of model bias that produce a bias dependence on the S\'ersic index $n$ as well as a different shear bias between galaxies consisting of a single S\'ersic profile and galaxies with a disc and a bulge. We also see an important coupling between several properties on the bias dependences. Because of this, we need to study several measured properties simultaneously in order to properly understand the nature of shear bias. This paper serves as a first step towards a companion paper that describes a machine learning approach to modelling shear bias as a complex function of many observed properties.
}


\maketitle

\section{Introduction}

Weak gravitational lensing is a powerful and  promising probe of cosmology for current and upcoming galaxy surveys, such as the Hyper Suprime-Cam (HSC; \citealt{Miyazaki2006}), the Dark Energy Survey (DES; \citealt{DES2005,Flaugher2005}), the Kilo Degree Survey (KIDS; \citealt{deJong2013}), the Large Synoptic Survey Telescope (LSST; \citealt{LSST2009}), Euclid \citep{Laureijs2011}, and the Wide-Field Infrared Survey Telescope (WFIRST; \citealt{Green2012}). Due to the gravitational potentials of the mass fluctuations between distant galaxies and us, the light is deflected, causing distortions in the images of the galaxies. By studying these distortions, we can infer and study the distribution of the  total matter (dark and baryonic) in the Universe. However, most of the galaxies are only distorted by a few percent of their intrinsic ellipticity value. Because of this, the ellipticity of the image of a galaxy is dominated by its intrinsic ellipticity, so we cannot measure the shear distortion of individual galaxies; instead, we can study them statistically if we have a sample of galaxies that is large enough so that the intrinsic ellipticities average out.

Cosmic shear measured from statistics on galaxy ellipticities allows us to improve the estimation of cosmological parameters by adding this new probe to cosmological studies. However, systematics in shear measurement propagate to cosmological parameter estimates. There are several systematics that make this measurement challenging \citep{Bridle2009}. Firstly, images are blurred due to the atmosphere or instrument response and suffer from other effects from the telescope optics.  Moreover, the convolution kernel of the image (point-spread function, or PSF) is not necessarily isotropic, varies spatially, and has to be estimated from either modelling or from the images of the stars from the same field. Secondly, the output images are pixelated. Finally, the pixels can suffer from Poisson noise coming from galaxy photons and other noise contributions such as sky background. Besides taking into account all these steps, we also need an accurate algorithm to estimate the galaxy ellipticities from the pixelated images.

All these effects can produce a bias on the estimation of the shear that can affect our statistics and cosmological analysis, and hence it is crucial to understand the nature of this bias in order to be able to either calibrate it or to improve our methodology to reduce its impact. Because of this, many studies have focused on the different sources of shear bias and calibration techniques. The shear bias is usually defined as multiplicative and additive factors that define a linear relation between the true and the measured shear.


One of the most studied sources of bias is noise, commonly referred to as noise bias \citep{Bridle2009,Bridle2010,Kitching2010,Kitching2012,Kitching2013,Refregier2012,Kacprzak2012,Melchior2012,Taylor2016}. \cite{Refregier2012}  present an analytic derivation for the bias of maximum likelihood estimators (MLEs) affected by an additive noise. They explore a simplified case where galaxy images are modelled and fitted with a Gaussian with its size as the single free parameter; they find a significant effect even for this simple approximation.  
\cite{Taylor2016} and \cite{Hall2017} present analytic descriptions of the impact of different sources of bias to dark energy measurements, finding noise bias to be the most relevant. They also present an analytic calibration of part of the bias. However, these expressions do not account for the full complexity of real images, and their precision is thus limited.

Other studies have shown many other potential sources of bias. Some examples are:
scale-dependence of bias on different cosmological parameters and redshifts \citep{Huterer2006,Amara2008,Kitching2015};
model bias coming from the assumptions of wrong models of galaxy morphology \citep{Massey2007,Voigt2010,Bernstein2010,Zhang2011,Kacprzak2012,Kacprzak2014,Mandelbaum2015};
selection bias coming from the fact that different samples of galaxies are affected differently by all these systematics  \citep{Kacprzak2012,Kacprzak2014};
limitations of model-fitting methods \citep{Melchior2010,Voigt2010} and how to improve them \citep{Bernstein2010};
galaxy morphology or size \citep{Mandelbaum2015,Clampitt2017};
PSF modelling and instrumental effects that cannot be treated as convolutions \citep{Massey2013};
the number of pixels in the PSF and the pixel integration level \citep{Voigt2010};
and the bulge-to-total flux ratio \citep{Voigt2010}.
\cite{Hoekstra2015} and \cite{Hoekstra2017} explored the sensitivity of multiplicative bias to the input parameters of simulated images and inferred the accuracy to which we need to measure the sizes and intrinsic ellipticities of galaxies for Euclid-like surveys.
Recently, \cite{Martinet2019} studied the impact of undetected galaxies in the image background on the shear calibration.

Finally, different shape estimators can lead to different biases and accuracies of the shear measurements. In order to compare a wide variety of estimators, several image processing challenges to put together different algorithms to estimate the shape of galaxies in the same set of simulations have been organized. The first challenges, known as the Shear Testing Programme STEP1 \citep{Heymans2006} and STEP2 \citep{Massey2007b}, showed the complexity of the shear measurement and the important role of shear bias. In order to improve the clarity in these studies,
the GREAT08 Challenge \citep{Bridle2009,Bridle2010} focused on a simplification of the problem, using a known PSF, simple galaxy models, and a constant shear. Later, in the
GREAT10 Challenge \citep{Kitching2010,Kitching2012,Kitching2013}, the realism was increased to include more complex galaxy morphologies, a varying applied gravitational shear, and some telescope systematics.
Finally, in the GREAT3 Challenge \citep{Mandelbaum2015},  different shape measurement methods were tested to infer weak lensing shear distortions from different simulated surveys (space- and ground-based), shear variations (constant or cosmologically varying), and galaxy morphologies (realistic and parametric).
They also studied the bias dependencies on truncation due to finite postage stamps, the S\'ersic index of the galaxy profiles, the PSF size, ellipticity and defocus, and the impact of the estimation and interpolation of the PSF. An encouraging conclusion of the study is that several methods were able to measure shear with systematic errors around the level required by Stage IV galaxy surveys. However, we note that GREAT3 had low sensitivity to noise bias due to the limited number of galaxies involved and the high signal-to-noise ratio (SNR) per galaxy.

In this paper we present a complementary study of the bias dependencies found in galaxy image simulations based on GREAT3 for different shape estimators. Our goal is to identify the main dependencies of bias found as a function of all simulation (input) and measured (output) parameters, PSF anisotropy, noise, pixelization, and model bias coming from the use of different shear estimators. This identification will be used as a first analysis (Paper I) and as motivation to develop a shear calibration method based on machine learning techniques in \cite{Pujol2020}, hereafter Paper II.
 In particular, we study the bias dependencies on all input parameters of the simulations and all output parameters obtained from the shear estimators in order to identify the properties to which bias is most sensitive.
 We also show some examples of the differences between ellipticity bias, which describes the errors on the estimation of the shape of the images, and shear bias, which defines the errors obtained in the estimation of the shear of a given sample of galaxies.
We study the method dependence by using two different and independent methods to estimate the shape and shear. One of the methods, \gfit\ \citep{Gentile2012,Mandelbaum2015}, is an MLE that measures the galaxy shape from fitting the best parameters from a given model. The second method is the Kaiser-Squires-Broadhurst (\cite{Kaiser1995}, hereafter \ksb) implementation of the public code \shapelens\ \citep{Viola2011}, which estimates the shape of the galaxy from the measurement of the weighted moments of the image.
We  also studied the effect of isotropic and realistic PSFs, noise, and pixelization by repeating the measurements of the estimators on new realizations of the image simulations; here, we applied some variations on the noise variance and the use of either an isotropic Gaussian PSF or different realistic and anisotropic PSFs. We do not explore the dependencies of bias coming from implementation parameters, such as the minimization or initialization parameters or the choice of different galaxy models, in this paper. However, given the agreement found between \gfit\ and \ksb\ on most of the dependencies, we think that the implementation of these methods does not significantly affect  the conclusions of this paper.

The paper is organized as follows. In Sect. \ref{sec:method}, we describe the image simulations, the shape estimators, and the methodology used to measure the shear and ellipticity bias. In Sect. \ref{sec:results}, we show and discuss the results of the main bias dependencies on input parameters (from the simulated images) and output parameters obtained from both estimators. We end in Sect. \ref{sec:conclusions} with a summary and discussion of the most important results of the paper. 

\section{Data and methodology}\label{sec:method}

\subsection{Images}

\begin{figure}
\centering
\includegraphics[scale=.6]{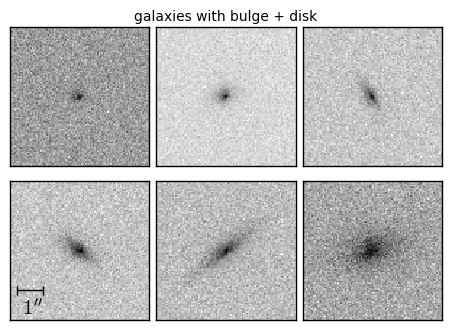}
\includegraphics[scale=.6]{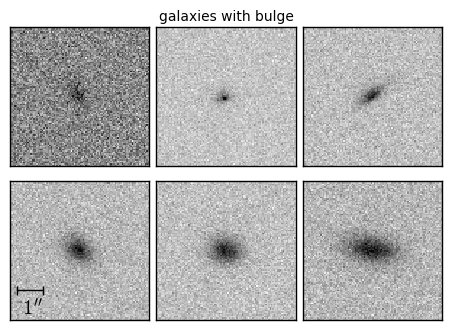}
\includegraphics[scale=.6]{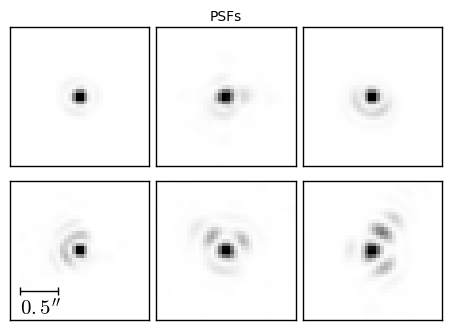}
\caption{Examples of galaxy and PSF images generated from \galsim. The top two rows show examples of galaxies with a de Vaucouleurs bulge and an exponential disc. The middle two rows show examples of galaxies with a single S\'ersic profile. In these two cases, we show galaxies of a variety of sizes and in increasing order. The bottom two rows show examples of PSF images, showing a wide range of complexities from simple and isotropic to complex and anisotropic. The PSF images have been zoomed in for visual reasons.}
\label{fig:gal_images}
\end{figure}

We used \galsim\ \citep{Rowe2015} to simulate the galaxy images of this analysis. We generated the images from the configuration parameters from the GREAT3 \citep{Mandelbaum2015} control-space-constant (CSC) branch for most of the study together with the centred corresponding PSFs. With this, we obtained images of $2\times 10^6$ galaxies corresponding to the GREAT3 CSC branch and their respective PSFs, with which we run our shear estimators. Each of the 200 images contains $100\times100$  stamps of $96$ pixels per side and a pixel scale of 0.05 arcsec with one galaxy in each stamp, giving a total of $10,000$ galaxies per image and a total of $2,000,000$ galaxies.
 In order to have an average intrinsic ellipticity of $0$ without the need for simulating more images, all galaxies have a  $90$-degree rotated counterpart. This was already the case for the GREAT3 Challenge. 
In every measurement of bias presented in this paper, we always included the orthogonal pairs of galaxies, or, if not, we corrected for the non-zero average ellipticity, as discussed later.
Two types of galaxies are included in the CSC branch: galaxies with a bulge using a single S\'ersic profile with a varying index $n$; and galaxies with a bulge defined from a de Vaucouleurs profile  and an exponential disc.
In Fig. \ref{fig:gal_images}, we show some examples of images of both types of galaxies, with the exponential disc (top two rows) and without the disc (middle two rows).
As in the GREAT3 CSC branch, we used $200$ different shear values and optical PSFs from a random distribution of values in a radius of $|g| < 0.05$, each of them assigned to each image of $10,000$ galaxies. Some examples of PSFs are shown in the bottom rows of Fig. \ref{fig:gal_images}. The distribution of the galaxies' morphological parameters (flux, bulge and disc profiles, radius, etc.) were obtained from fits on real Hubble Space Telescope (HST) images from the COSMOS F815W<23.5 sample so that they represent a realistic distribution. The orientation angle is set randomly, including an orthogonal version of each galaxy, cancelling shape noise but bringing a random distribution of orientation angles. The mean and variance of background noise is also estimated from the real images.  More details on the parameters and characteristics of the simulations are available in the GREAT3 Challenge Handbook \citep{Mandelbaum2014}, where the authors give a full description of the fitting process and the properties of the galaxy catalogue, the instrument parameters, and the image simulations.

Additionally, in order to study effects such as truncation, miscentring, or PSF effects, we also generated simulations with small variations with respect to the original ones corresponding to the GREAT3 CSC branch described above. In particular, we generated the following simulated images. On one side, we forced all the images to be well centred in the stamps. As the \cgfit\ minimizer used allows for the possibility of leaving specific galaxy model parameters fixed while fitting, we used this feature to fix the centre positions of the galaxies to the correct ones in order to study miscentring. Comparing these simulations with the previous ones, we can measure the effects of miscentring on shear and ellipticity bias. Secondly, instead of using the original PSFs from GREAT3 CSC, we used a Gaussian isotropic PSF to generate the images. This allows us to evaluate the impact of the PSF anisotropies on the bias measurements. Finally, we generated other simulations where we reduced or increased the variance of the images' Gaussian noise. In particular, we generated simulations applying a factor of $4$ or a factor of $1/4$ to the noise variance from GREAT3. We also repeated this for the images with smaller pixels described above in order to see the correlation between pixel size and noise on the effects of ellipticity and shear bias.
These additional image simulations are used for Sect.~\ref{sec:other_tests}, so the rest of the results are based on the fiducial set of simulations.

\subsection{Image processing}

We used two different shape estimators to measure shear and ellipticity. We then compared the different results to see how much our study depends on the estimators used. Below we describe the two estimators.

\subsubsection{\cgfit}

The \cgfit\ method \citep{Gentile2012,Mandelbaum2015} is a maximum-likelihood shape estimator. A forward model fitting algorithm is used to minimize an $\chi^2$ between the simulated patch and a parametric model generated using \galsim. We chose to use the native minimization algorithm provided by \cgfit, which is based on cyclic coordinate descent.

The chosen model  \citep[the same one used in][]{Mandelbaum2015} implements galaxy profiles as a weighted sum of an exponential disc and a de Vaucouleurs bulge, with eight parameters: centroid position, ellipticity, flux, flux ratio, and half-light radius, each for bulge and disc . We ran \pse\ \citep{Bertin1996} to initialize the centroid estimates. The method estimates the ellipticity defined as:

\begin{equation}
\epsilon = \epsilon_1 + i \epsilon_2 = \frac{Q_{11} - Q_{22} + 2i Q_{12}}{Q_{11} + Q_{22} + 2\sqrt{Q_{11}Q_{22} - Q^2_{12}}},
\label{eq:ellip}
\end{equation}
with
\begin{equation}
Q_{ij} = \frac{\int{d ^2 x I({\bf x}) W({\bf x}) x_ix_j} }{\int { d^2 x I({\bf x}) W({\bf x}) }},
\end{equation}
where $I({\bf x})$ is the surface brightness at the image position $\bf x$ and $W({\bf x})$ is a weighting function used to suppress the noise contribution at large scales. We note that $Q_{ij}$ is not used for the model fitting method and that it is only introduced to define the ellipticity parameters that are measured.

It is important to note that our simulations are built from either a single S\'ersic model or from the weighted sum of an exponential and a de Vaucouleurs profile with different ellipticity and orientation, contrary to the model used in the fitting. All these factors can result in significant model bias in the estimation of galaxy shapes. 


\subsubsection{\shapelens}

This public \textsc{C}++ library includes several modules to estimate the shape of galaxy images. One of them is presented in
\cite{Viola2011} and is based on the KSB method \citep{Kaiser1995}. This method estimates the shape from the surface-brightness moments of the images according to the following definition of ellipticity:
\begin{equation}
\chi = \frac{Q_{11} - Q_{22} + 2i Q_{12}}{Q_{11} + Q_{22} }.
\end{equation}
Although the ellipticity definition from Eq. (\ref{eq:ellip}) is supposed to be more precise, it is also noisier and, because of this, is not used for the implementation of this model. We note that the two estimators use different ellipticity definitions with a different ellipticity modulus. The comparison between the two methods in the paper shows how these different algorithms and definitions affect shear bias when average measurements are made.

To compute these moments, the method uses an isotropic weighting function, whose size depends on the estimation of the galaxy size. Due to the isotropy of the weighting function, this estimation produces a bias that increases with the ellipticity. However, this effect can be corrected by considering the higher order contributions that the weighting function makes to the shape measurements, as discussed in \cite{Viola2011}. This correction can be directly implemented from \shapelens. From all the implementation modes available in \shapelens, we implemented the one that uses the trace of the first order correction \citep[Eq. (33) from][]{Viola2011}, since it gives the best results. Throughout the paper we refer to this implementation as \ksb.

 By design, this estimator does not involve any analytic form for the galaxy shape. However, the analysed pixels  are weighted with an isotropic Gaussian kernel from a pre-selected family of size, which can also produce a model bias.

 \subsection{Bias measurement}

We describe the relation between the observed ellipticities $\epsilon_{\rm obs}$ from our shape estimators and the true ellipticities $\epsilon$ (from both intrinsic shape $\epsilon_{\rm int}$ and shear $g$) as follows:

\begin{equation}
\epsilon_{i,\rm obs} = a_i + (1 + b_i) \epsilon_{i} = a_i + (1 + b_i)(\epsilon_{i,\rm int} + g_i),
\label{eq:b_relation}
\end{equation}
where $i = {1,2,+,\times,}$ and $a_{i},b_{i}$ are the additive and multiplicative ellipticity bias parameters and describe the errors produced on the estimation of the images' shapes with respect to their true shapes. We measured them from a linear fit to the scatter distribution between $(\epsilon_{i,\rm int} + g_{i})$ and $\epsilon_{i,\rm obs}$. The tangential and radial components $+$ and $\times$ refer to the alignment with respect to the PSF. 

As the mean intrinsic ellipticity of galaxy samples is zero and its shear $g$ is constant, we can describe the relation between the mean observed ellipticity and shear as:

\begin{equation}
\langle \epsilon_{i, \rm obs} \rangle =  g_{i,\rm obs} = c_i + (1 + m_i) g_i ,
\label{eq:g_relation}
\end{equation}
where $i = {1,2,+,\times,}$ and now $c_{i},m_{i}$ are the additive and multiplicative shear biases (see \cite{Huterer2006} and \cite{Heymans2006} for their Taylor expansion notation). Shear bias describes the sensitivity of the shape estimators to small distortions with respect to the intrinsic ellipticity. We note that ellipticity and shear bias describe different errors and sensitivities produced in the shape measurements, and thus their behaviours are not necessarily similar.

We measured these parameters in two steps. First, we measured $\langle \epsilon_{i,\rm obs}\rangle$ and its error $\sigma_{\langle \epsilon_{i,\rm obs}\rangle}$ for each set of galaxies with the same value of $g_{i}$. Second, with these measurements we linearly fitted  $g_{i,\rm{obs}}$ versus $g_{i}$ using  $\langle \epsilon_{i,\rm obs}\rangle$ and weighted by $1/\sigma_{\langle \epsilon_{i,\rm obs}\rangle}$, as estimated in the first step. We calculated $\sigma_{\langle \epsilon_{i,\rm obs}\rangle}$ by performing a jackknife (JK) method  on $50$  balanced subsamples. We checked that the errors obtained when using more than $20$ subsamples did not depend on the number of subsamples used. We also checked that the distribution of the results of the JK subsamples is consistent with a Gaussian distribution and did not find outliers in these distributions, which suggests that the error estimation used here is sufficiently describing the scatter in the results.
We excluded from the analyses galaxies whose shape is wrongly estimated (with a failed measurement or an ellipticity modulus larger than 1) as well as their orthogonal pairs. The final catalogue for the two methods is slightly different because the rejected objects  and their numbers are not the same. This does not have a significant impact on the results of the paper, with the exception of small differences in the error bars.


Depending on the properties used to define our galaxy samples (in particular when using output properties), we find situations where the mean intrinsic ellipticity is not consistent with $0$. In these cases, the estimated parameters from these formulas are very sensitive to the residual ellipticities. This can be taken into account using the following estimators for $c$ and $m$:

\begin{equation}
g_{i,\rm obs}  = c_i + (1 + m_i) (\langle \epsilon_{i, \rm int} \rangle + g_{i})
\label{eq:g_relation_meanint_corr}
\end{equation}
and then again computing the mean ellipticities over the different values of $g_{i}$.
These formulas are the equivalent to Eq. (\ref{eq:g_relation}) when $\langle \epsilon_{i, \rm int} \rangle = 0$. When this was not the case, we used this formula in order to compensate for the effects of the residual of $g_{i, \rm int}$ on $c$ and $m$. We have found this to happen with a similar frequency for both shape estimators.


\begin{figure*}
\centering
\includegraphics[width=.45\linewidth]{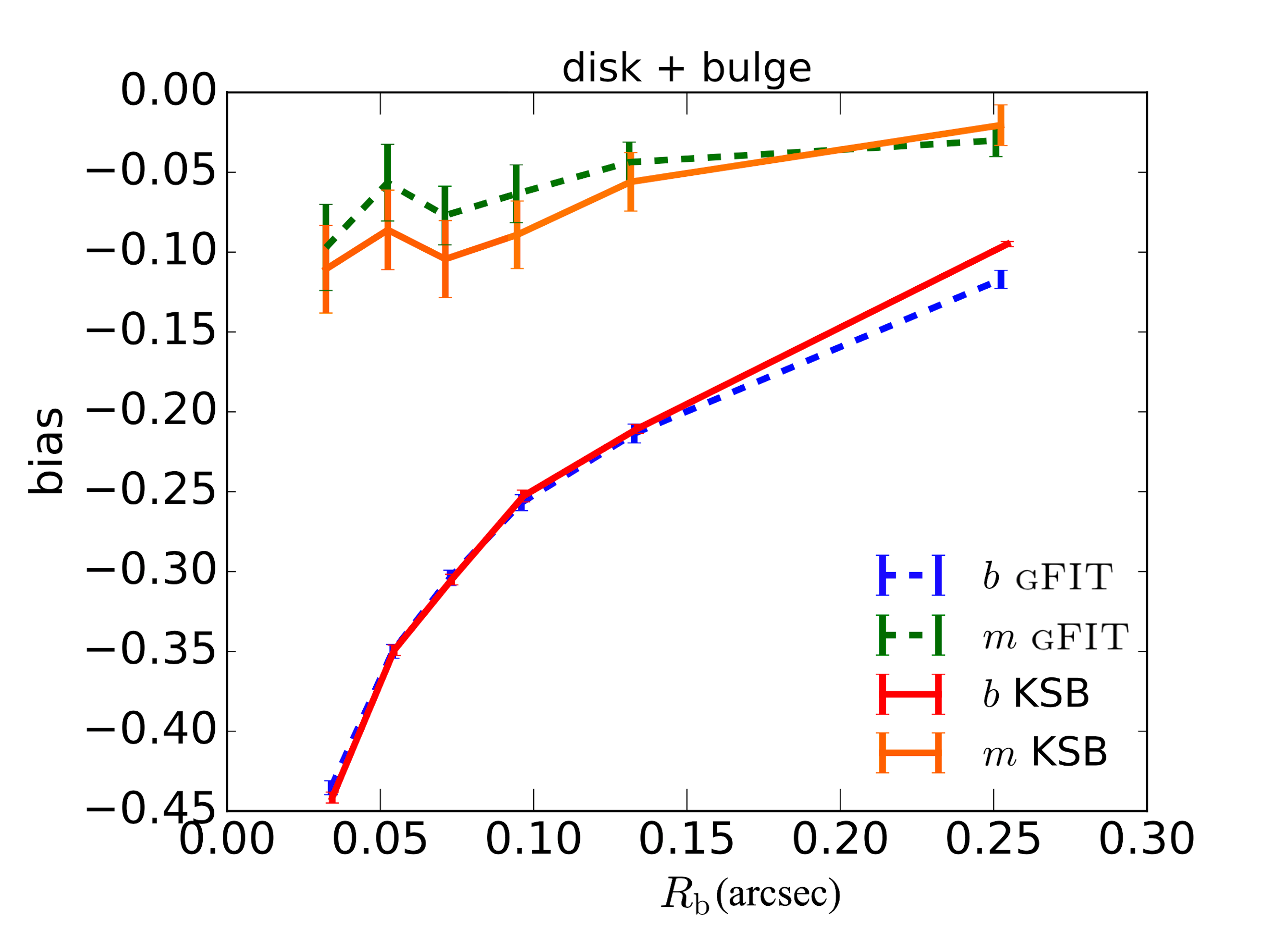}
\includegraphics[width=.45\linewidth]{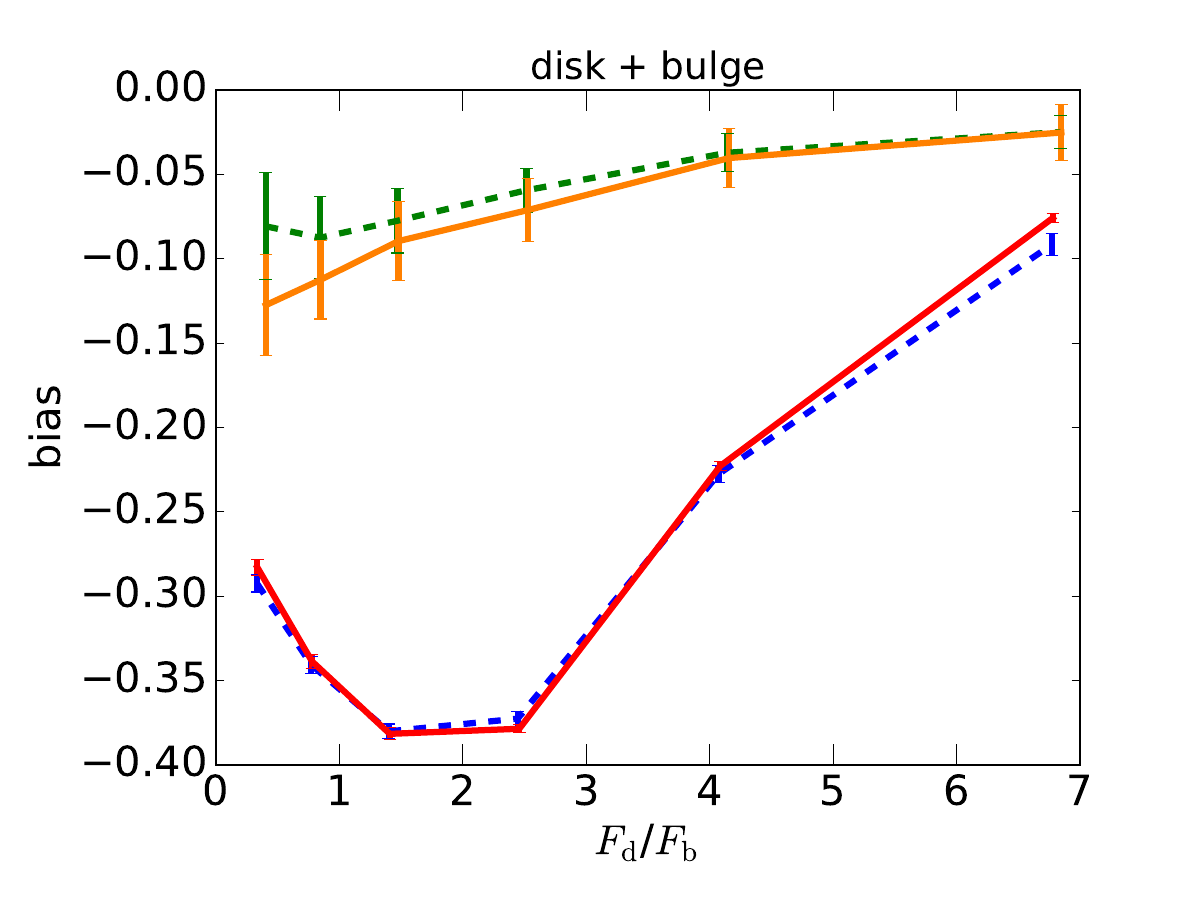}
\includegraphics[width=.48\linewidth]{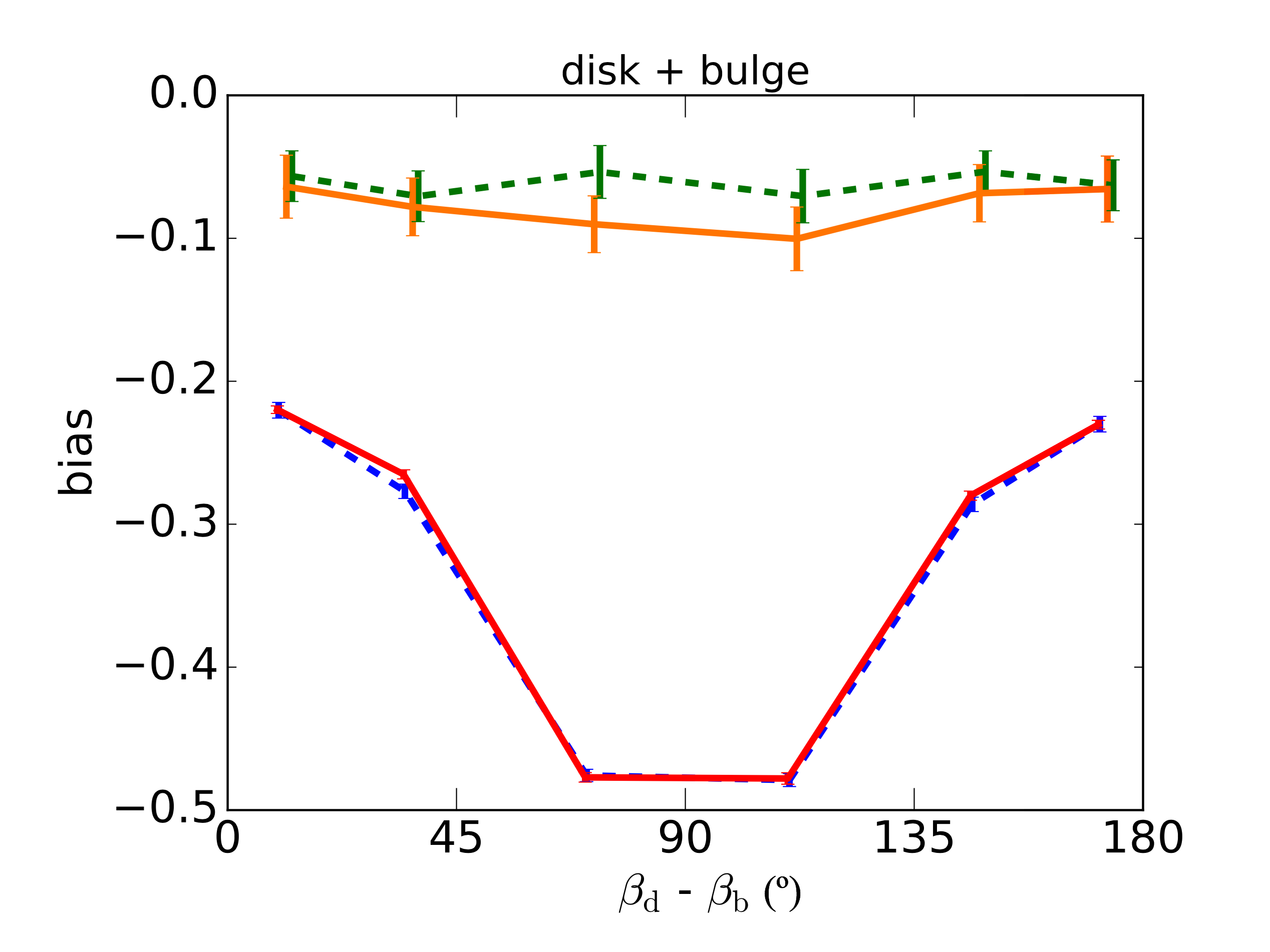}
\includegraphics[width=.45\linewidth]{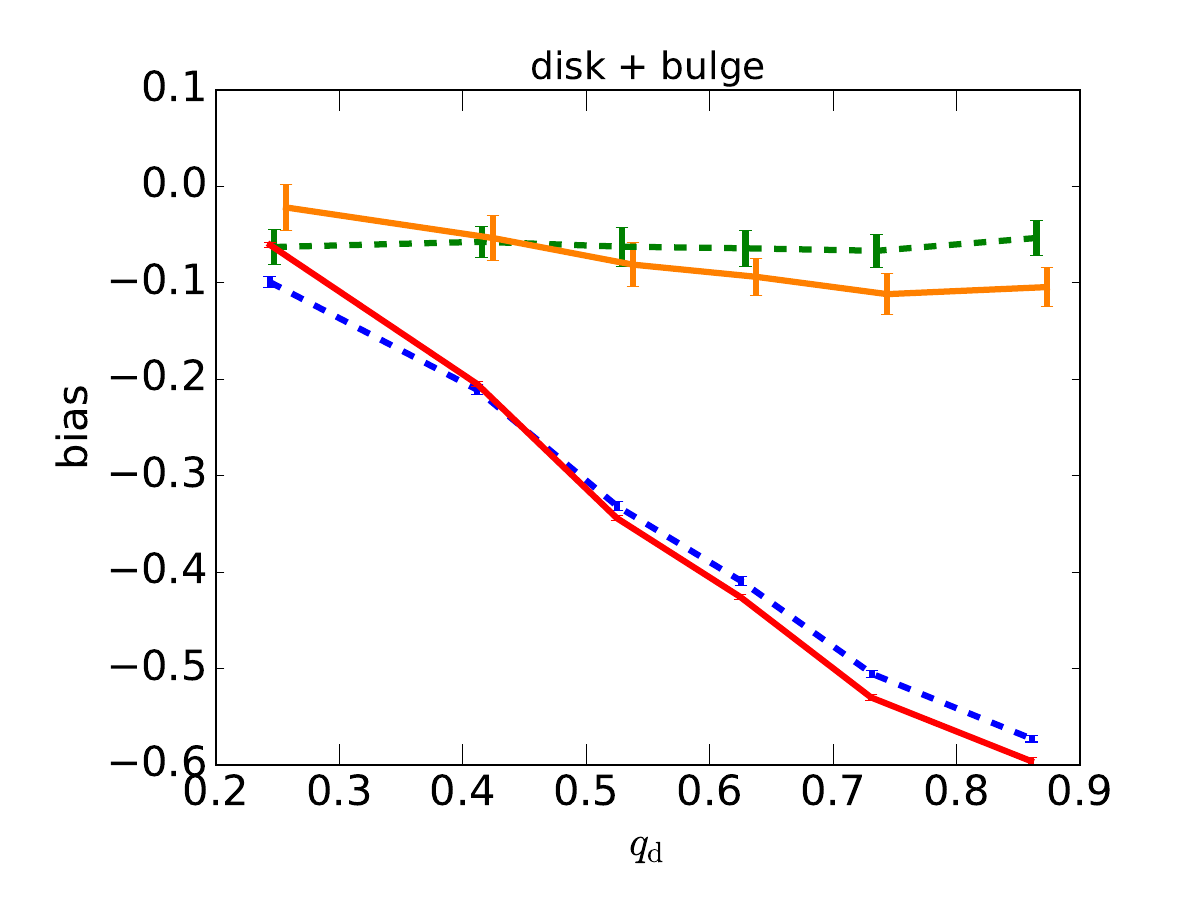}
\caption{Multiplicative ellipticity bias ($b_{1,2}$ from Eq. (\ref{eq:b_relation})) and shear bias ($m_{1,2}$ from Eq. (\ref{eq:g_relation})) as a function of the input galaxy properties, such as bulge size $R_{\rm b}$ (top left),  disc-to-bulge flux ratio (top right), the difference between the disc and bulge orientation angles (bottom left), and disc axis ratio $q$ (bottom right) for galaxies with a disc and a bulge and for both models \cgfit\ and \ksb. }
\label{fig:bm_plots_size}
\end{figure*}

\section{Results}\label{sec:results}

We studied the ellipticity and shear bias dependencies on all the input properties available from the image catalogues generated from \galsim\ (i.e. the grand-truth parameters that describe the galaxies and characteristics of the simulated images) and on all the output parameters obtained from both \ksb\ and \cgfit. We noticed a few parameters that strongly impact the bias. Moreover, we find that ellipticity bias is sensitive to more properties than just shear bias.
In this section, we focus on the main properties that strongly impact ellipticity and shear bias.
 For all the properties and bias measurements shown in the paper, the applied bins  are defined so that each bin contains the same amount of galaxies, and the biases are shown as the mean of the two components 1 and 2 unless specified.

\subsection{Ellipticity bias versus shear bias}


In \Fref{fig:bm_plots_size}, we show some examples of different behaviours between ellipticity and shear bias.
In many cases, shear bias $m$ is very different than ellipticity bias $b$; the difference comes not only from the amplitude but also from the behaviour of the dependencies of bias. These differences illustrate the different concepts behind both biases mentioned in Sect. \ref{sec:method}.
A large ellipticity bias does not imply a large shear bias because even if our estimator does not correctly predict the ellipticity of an image, it could still correctly capture small changes around this ellipticity.
We can see that $b$ is generally significantly below $0$, with an average value of around $-0.25$ for the galaxies with a bulge and a disc. On the other hand, $m$ tends to be much more consistent with $0$, having an average value of approximately $-0.05$. This indicates that, although we do not recover the correct individual ellipticities of the galaxies when they have a bulge and a disc (and thus have a large ellipticity bias), we still detect the shear signal from shear (that is, we have a low shear bias).
This is consistent with previous analyses, which established that shear bias has to be considered over sample statistics rather than from the individual shape measurements, since individual ellipticities (and their biases) are not meaningful for cosmological analysis \citep{Bernstein2002,Bernstein2014,Israel2016,Conti2016,Huff2017,Mandelbaum2017}.
For this reason, we focus on shear bias dependences in the rest of the paper.

We also note the consistency between the two \ksb\ and \cgfit\ estimators. The agreement indicates that, at least at the precision level of this study and for these image simulations, both methods respond similarly to the simulated image characteristics.

\begin{figure*}
\centering
\includegraphics[width=.48\linewidth]{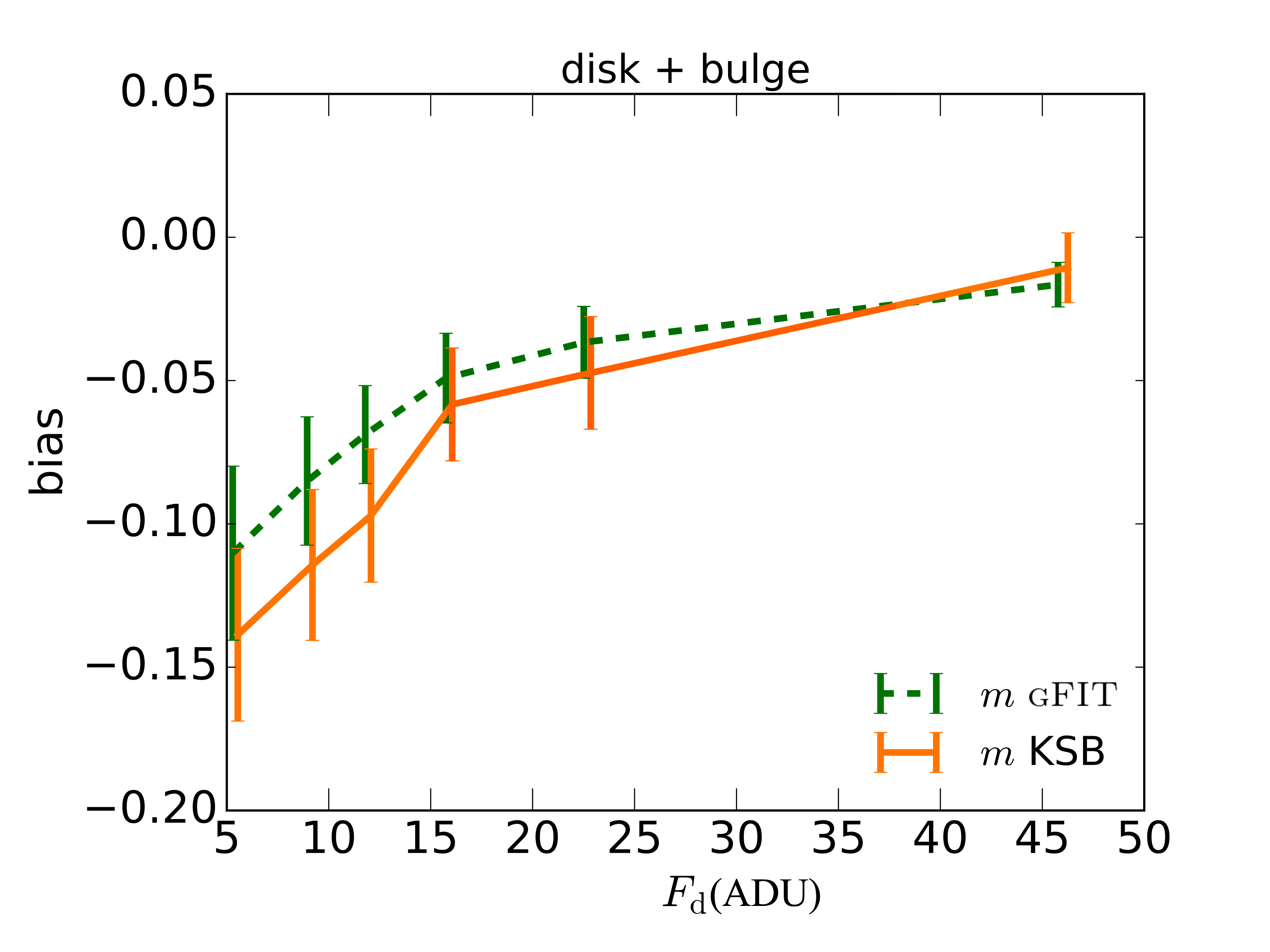}
\includegraphics[width=.48\linewidth]{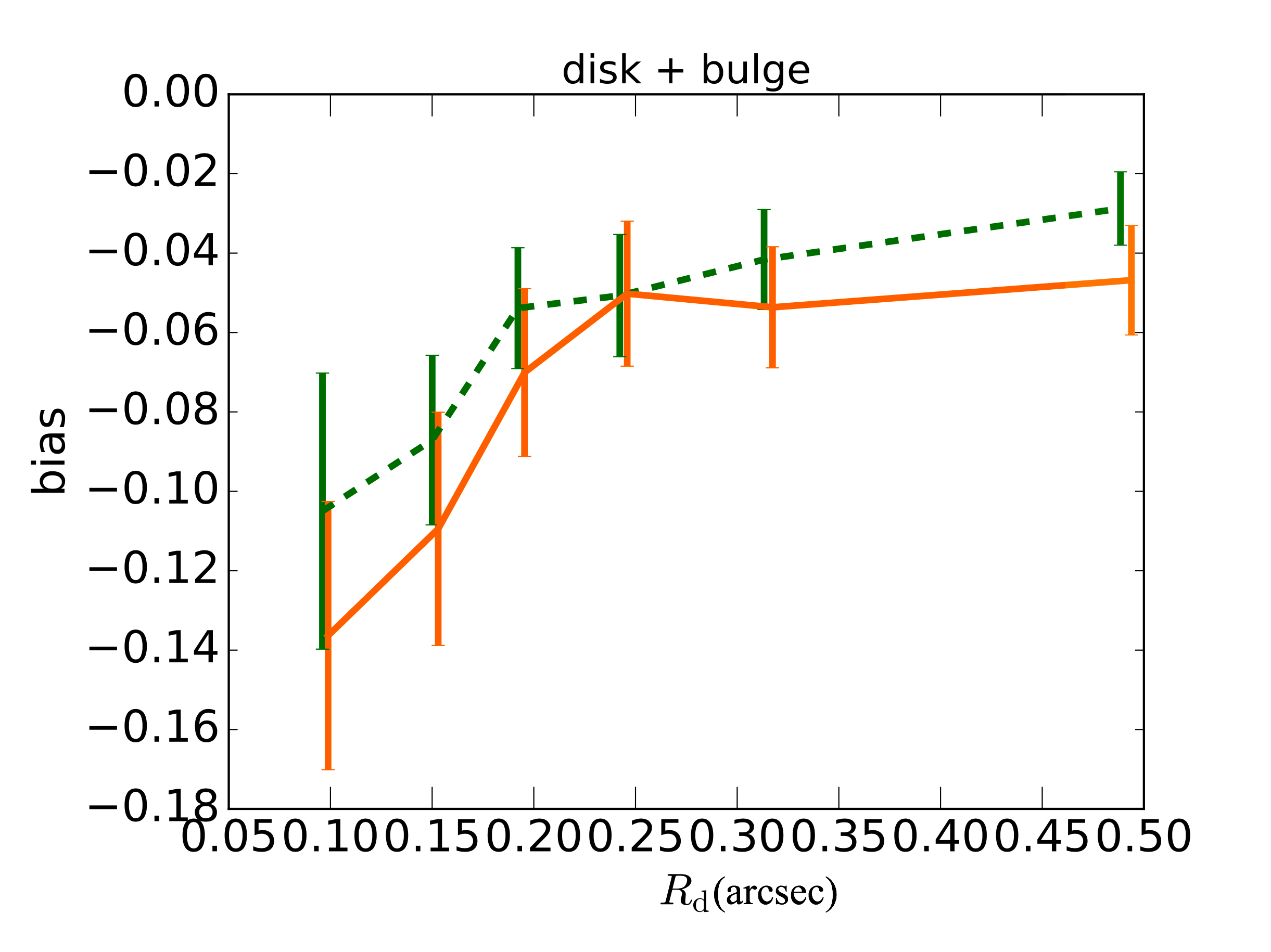}
\includegraphics[width=.48\linewidth]{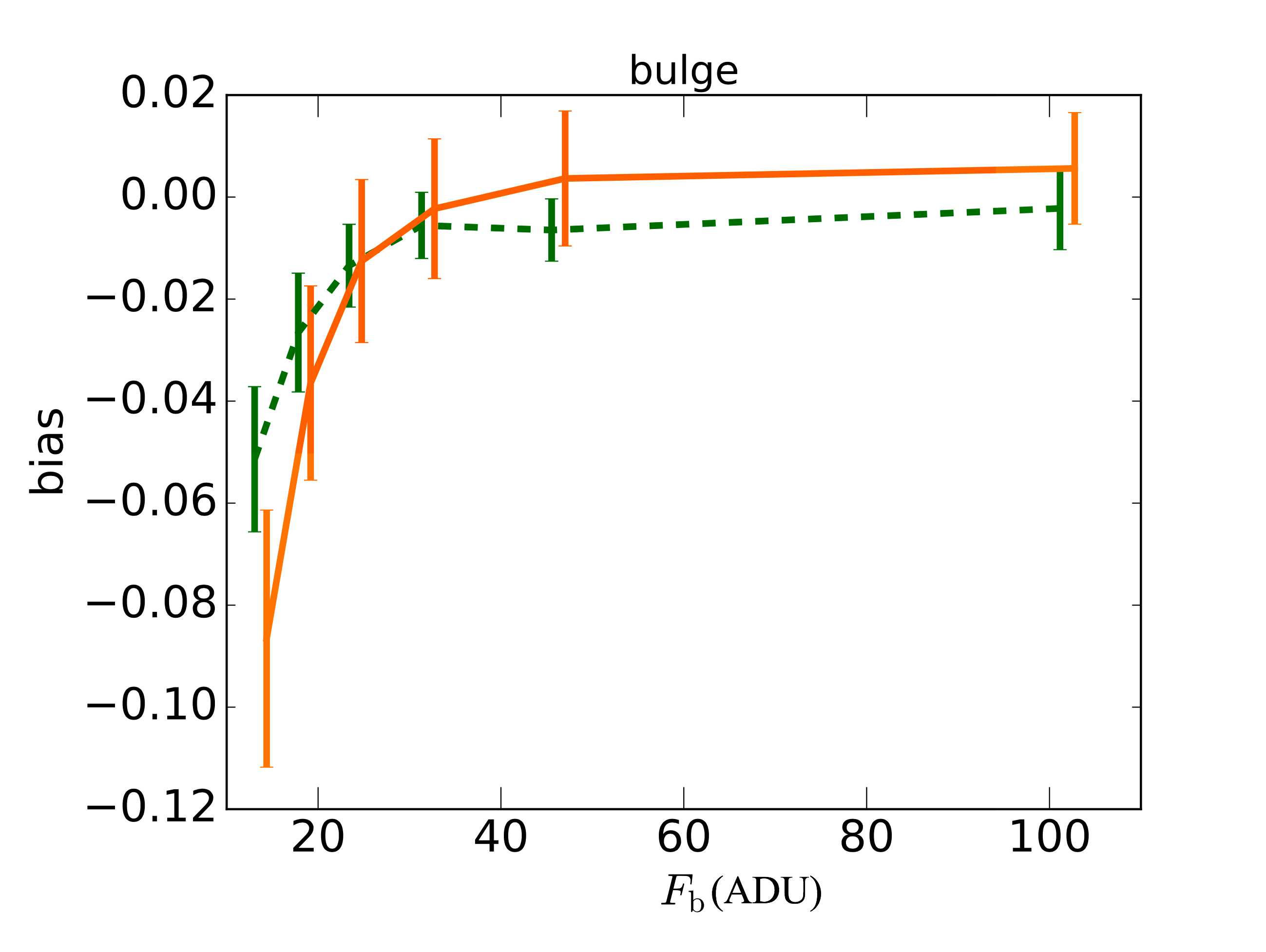}
\includegraphics[width=.48\linewidth]{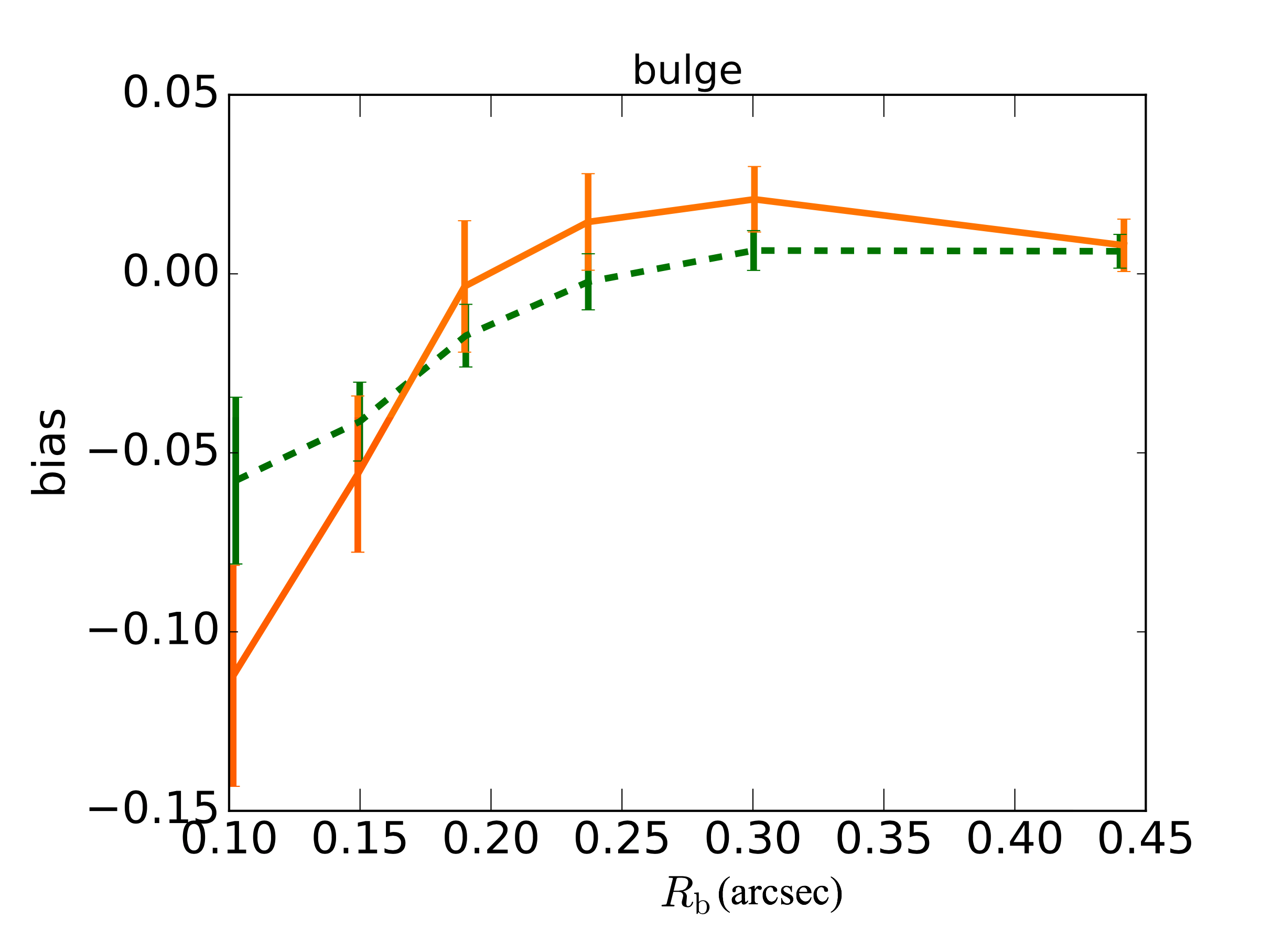}
\caption{Multiplicative shear bias as a function of the input disc flux (top left) and size (top right) for galaxies with a disc and a bulge, and as a function of bulge flux (bottom left) and size (bottom right) for galaxies with only a bulge. Green dashed lines show the results of the average $m$ for \cgfit,\ and the orange solid lines show $m$ for \ksb.}
\label{fig:in_plots_size}
\end{figure*}

\subsection{Shear bias dependencies on input parameters}\label{sec:shear_bias}

In Figs. \ref{fig:in_plots_size}, \ref{fig:in_plots_beta}, and \ref{fig:in_plots_model}, we show the input properties on which we found the strongest shear bias dependencies. Again, we note the good agreement between both shear estimators, \ksb\ and \cgfit,  given the precision of the errors of our analysis; they show similar behaviours for all the image characteristics (except for one case that we mention later).

\subsubsection{Effects of size, flux, and ellipticity}

In Fig. \ref{fig:in_plots_size}, we show the shear bias dependencies on size and flux. We note that in these cases the flux and size of the bulges correspond to the total flux and size of the galaxies; however, the disc information from the top panels does not give all the information about the size and flux of the object, since the bulge can be significant in some cases.

We see that shear bias tends to increase with galaxy flux and size. This indicates that the estimators give better results for larger images, as expected, since the signal of the image is better. Although the error bars are large, we can see a difference of around $0.1-0.15$ on the bias from the first to the last bins. In the case of galaxies without discs, shear bias is consistent with $0$ for bright and large galaxies.

\subsubsection{Effects of orientation and shape}

\begin{figure}
\centering
\includegraphics[width=.99\linewidth]{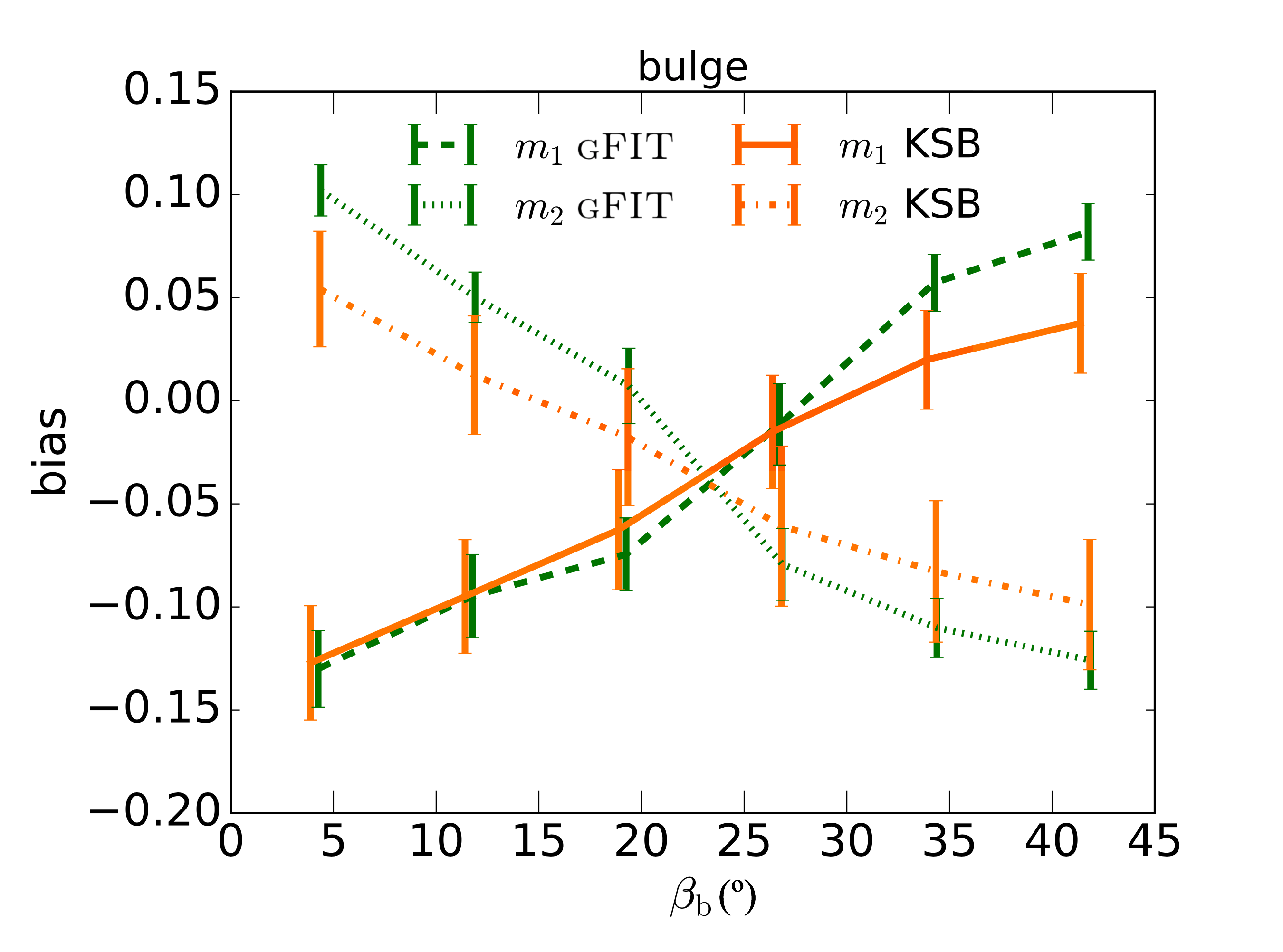}
\includegraphics[width=.99\linewidth]{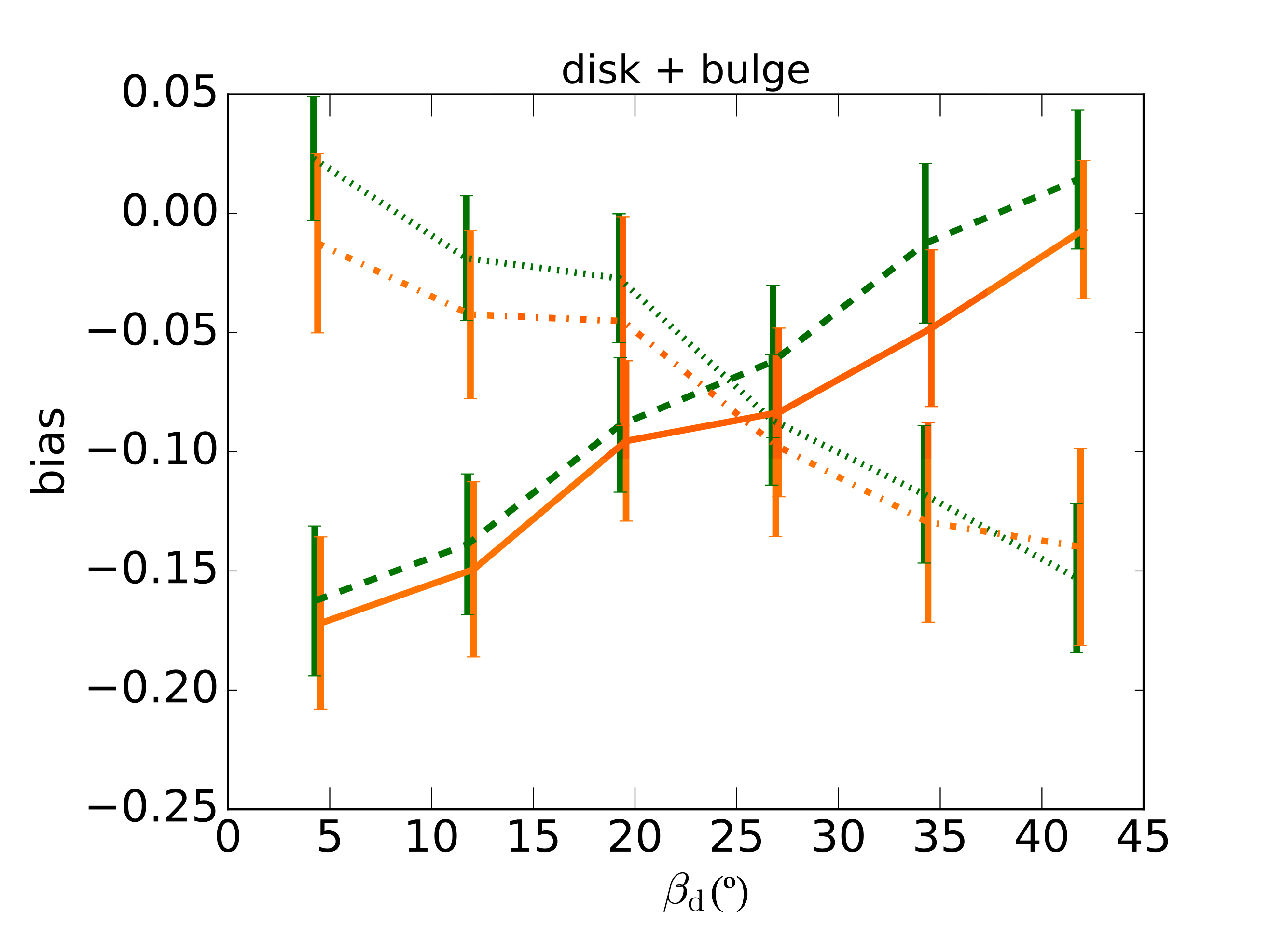}
\includegraphics[width=.99\linewidth]{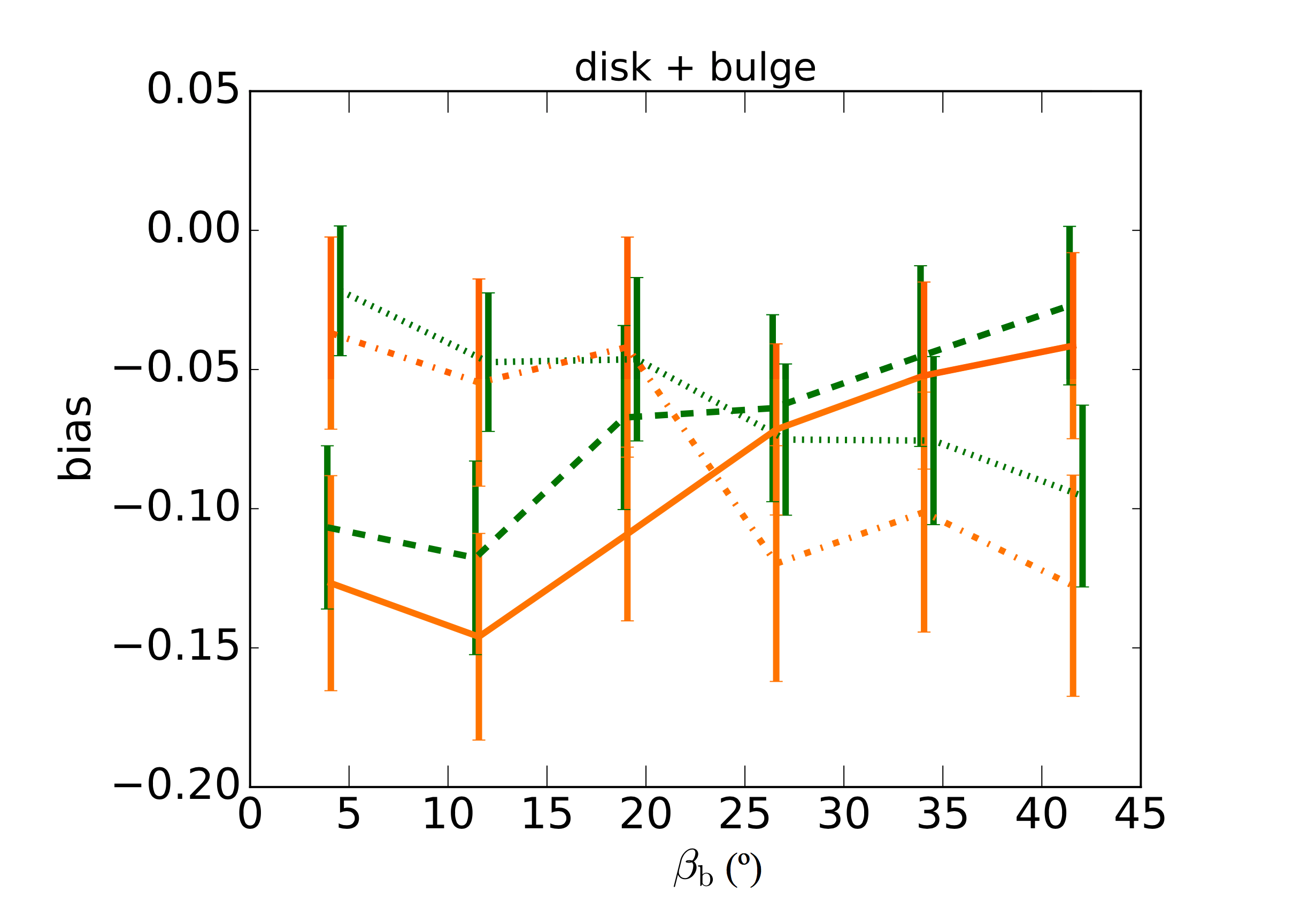}
\caption{Multiplicative shear bias as a function of the input orientation angle for galaxies with only a bulge (top), and as a function of the disc (middle) and bulge (bottom) orientations for galaxies with a bulge and a disc. Green lines show the results of $m_1$ (dashed) and $m_2$ (dotted) for \cgfit,\ and orange lines show $m_1$ (solid) and $m_2$ (dash-dotted) for \ksb. }
\label{fig:in_plots_beta}
\end{figure}

In Fig. \ref{fig:in_plots_beta}, we focus on the parameters related to the image orientation and we show the shear bias as a function of the bulge orientation angle $\beta_{\rm b}$ for galaxies with only a bulge, that is the global orientation of these images (top panel), the disc orientation angle $\beta_{\rm d}$ of the galaxies with a disc and a bulge (middle panel), and $\beta_{\rm b}$ for the same galaxies (bottom panel).

We see the same effect on $m$ in all the cases, although with different amplitudes, and $m_1$ and $m_2$ show antisymmetric dependencies. While $m_1$ increases (it has a positive slope) with $\beta$, $m_2$ decreases with a similar amplitude. The dependencies are symmetric with respect to $45$ degrees; this is the reason why we only show the range from $0$ to $45$ degrees. In order to have a zero mean ellipticity in all the bins, we included the orthogonal pairs of the galaxies in each bin so that the bins that have galaxies with orientation angle $\beta$ also include galaxies with orientation $\beta + 90^o$.

The fact that the dependence is weaker for the galaxies with a disc and a bulge (middle and  bottom panels) is expected, since the orientation angle of the two components can be different and hence the global orientation is less clear. This is because the disc and bulge parameters are fitted independently, forcing them to have the same centroid but allowing the orientation angle to be different. The half-light radius of the bulge is forced to be smaller than that of the radius, and the bulge-to-total flux ratio is constrained to be between 0.1 and 0.9 (otherwise a single Sérsic profile is fitted). As a result of this, galaxies with the two components have a wide range of bulge-to-disc radii and fluxes with non-negligible differences between the disc and bulge orientation angles. The fact that the weakest dependence is for the orientation angle of the bulge (bottom panel) suggests that the disc contribution to the global galaxy orientation is higher than the bulge contribution.

\begin{figure}
\centering
\includegraphics[width=.98\linewidth]{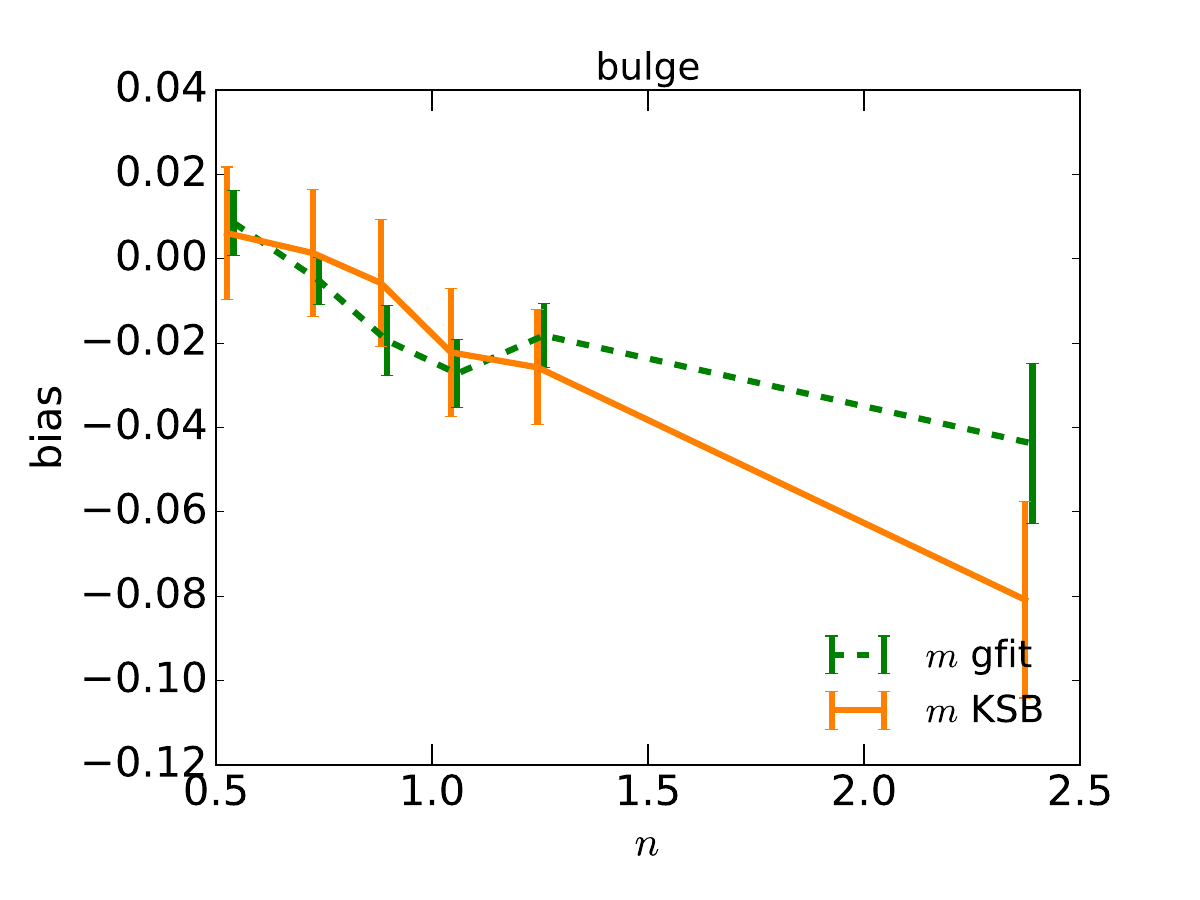}
\includegraphics[width=.98\linewidth]{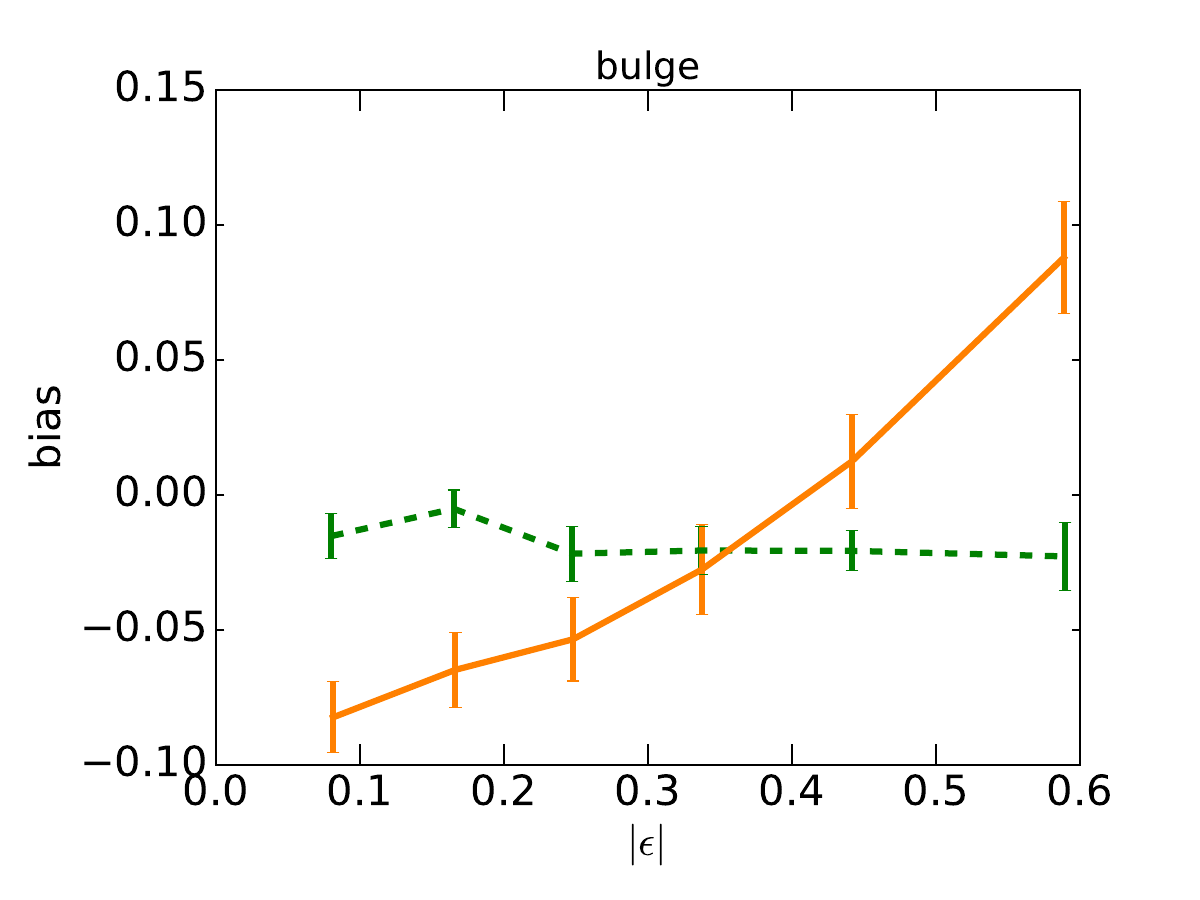}
\includegraphics[width=.98\linewidth]{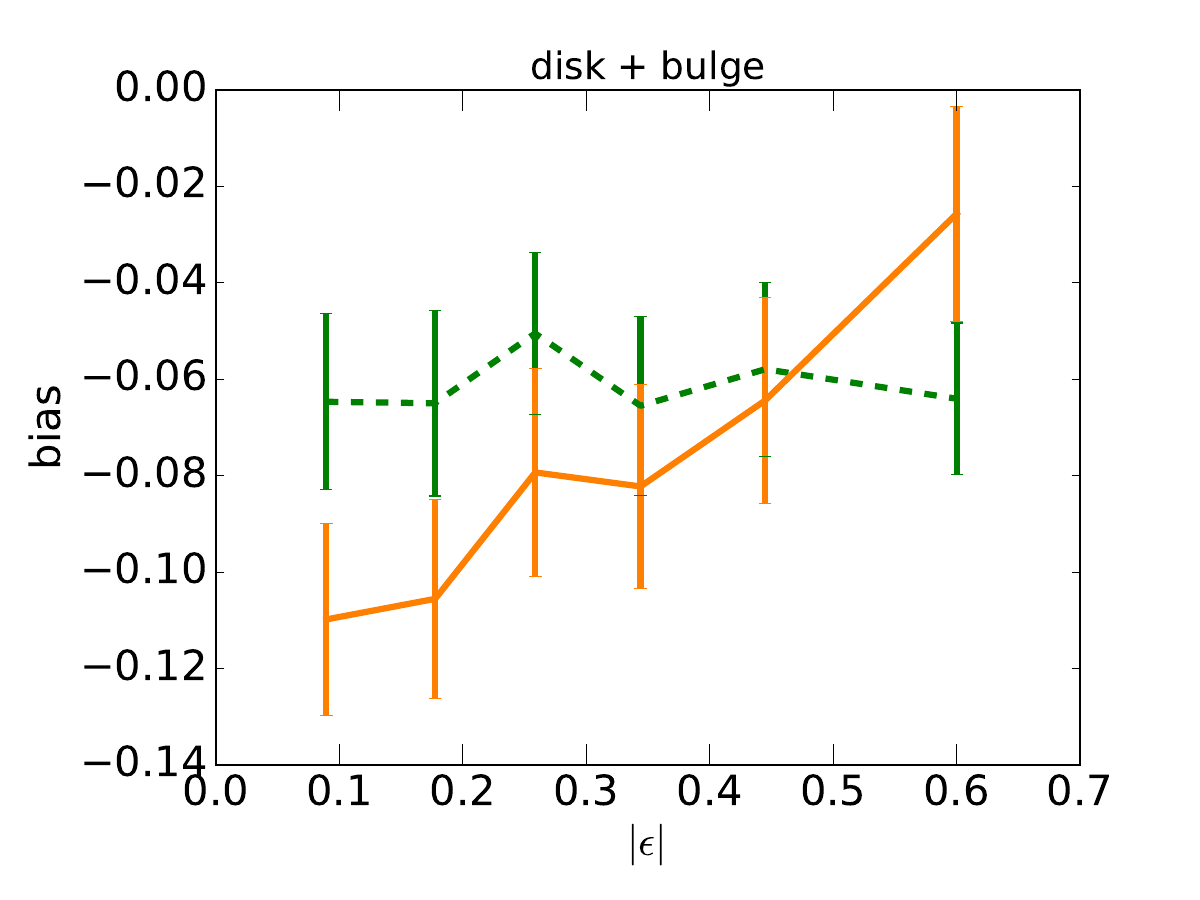}
\caption{Multiplicative shear bias as a function of the input S\'ersic index $n$ for galaxies with only a bulge (top), and as a function of the modulus of the intrinsic ellipticity $|\epsilon|$ for galaxies with only a bulge (middle) and with a bulge and a disc (bottom). Green dashed lines show the results of $m$  for \cgfit,\ and orange solid lines show $m$ for \ksb.}
\label{fig:in_plots_model}
\end{figure}

The shear bias seen in this figure can be explained from pixelization effects, since the bias is correlated with the pixel directions. Due to the direction of the pixels and its discretization, galaxies aligned with the pixels  (represented in the first bins) will affect the flux of the nearby pixels less if the image is sheared towards its direction than if the shear causes a rotation of the image, and hence $m_1$ will be more negative than $m_2$. Exactly the opposite happens when the images are oriented to the diagonal of the pixels (shown in the last bins), where small distortions of the image towards the diagonal of the pixels will impact the closest pixels less than small rotations. We find that the amplitude of the effect decreases with the size of the galaxy images, since in this case the proportion between the galaxy and the pixel size is larger. In particular, with the pixel scale and noise level corresponding to the GREAT3 images, we find that for galaxies with a disc half-light radius smaller than 0.1, the amplitude of the effect is approximately $20\%$, while for galaxies larger than a half-light radius of 0.4, the amplitude is approximately $10\%$.


Analysing shear bias dependencies on orientation angles has not been a priority for weak lensing studies because galaxy samples are never selected based on galaxy ellipticities. However, galaxy orientation can be important when the alignment of galaxies with the PSF or pixel grid has an effect on the selection and shear bias of the sample. The shear bias dependences on orientation need to be well understood for a better shear bias calibration. For this reason, in Paper II we incorporate the orientation angle as input data for the machine learning algorithm to improve the characterization and calibration of shear bias.

\subsubsection{Model bias}\label{sec:model_bias}

In this subsection, we show an example of model bias found in the study. We have already seen that model bias affects the two simulated galaxy populations  differently: $m$ is in general consistent with $0$ for galaxies without a disc but approximately $-0.05$ for galaxies with a disc. Although this is not a rigourous estimate of the impact of model bias on these shape estimators, it gives us an idea of how much different galaxy profiles can affect the shear estimation for these two estimators.

Another model bias example is shown in the top panel of Fig. \ref{fig:in_plots_model}, where we show the dependence of shear bias on the S\'ersic index $n$ for the galaxies with only a bulge. We have used the whole range of Sérsic index values, $0.1 < n < 6$, and distributed them into equally dense bins. The bins are located in the mean values of $n$ for each bin. Bias increases up to a $10\%$ bias for high S\'ersic index. This effect can come from two contributions. On one side, our fitted models do not include arbitrary  S\'ersic profiles, and this can cause a model bias. On the other hand, a large S\'ersic index $n$ corresponds to a steep decrease in luminosity, which makes the luminosity of these galaxies concentrated in the centre. Hence, these galaxies can be detected as small, occupying few pixels, which makes the estimation of the ellipticity and the interpretation of small distortions difficult. \cite{Kacprzak2014} also studied the shear bias dependence on the S\'ersic index, finding opposite trends (increasing with $n$ instead of decreasing). In their case they used an MLE with a different galaxy model than in \gfit; therefore, the difference between \cite{Kacprzak2014} and our results shows how sensitive the trend can be to different assumptions in the models used.

\subsection{Bias dependencies on output parameters}

\begin{figure*}
\centering
\includegraphics[width=.45\linewidth]{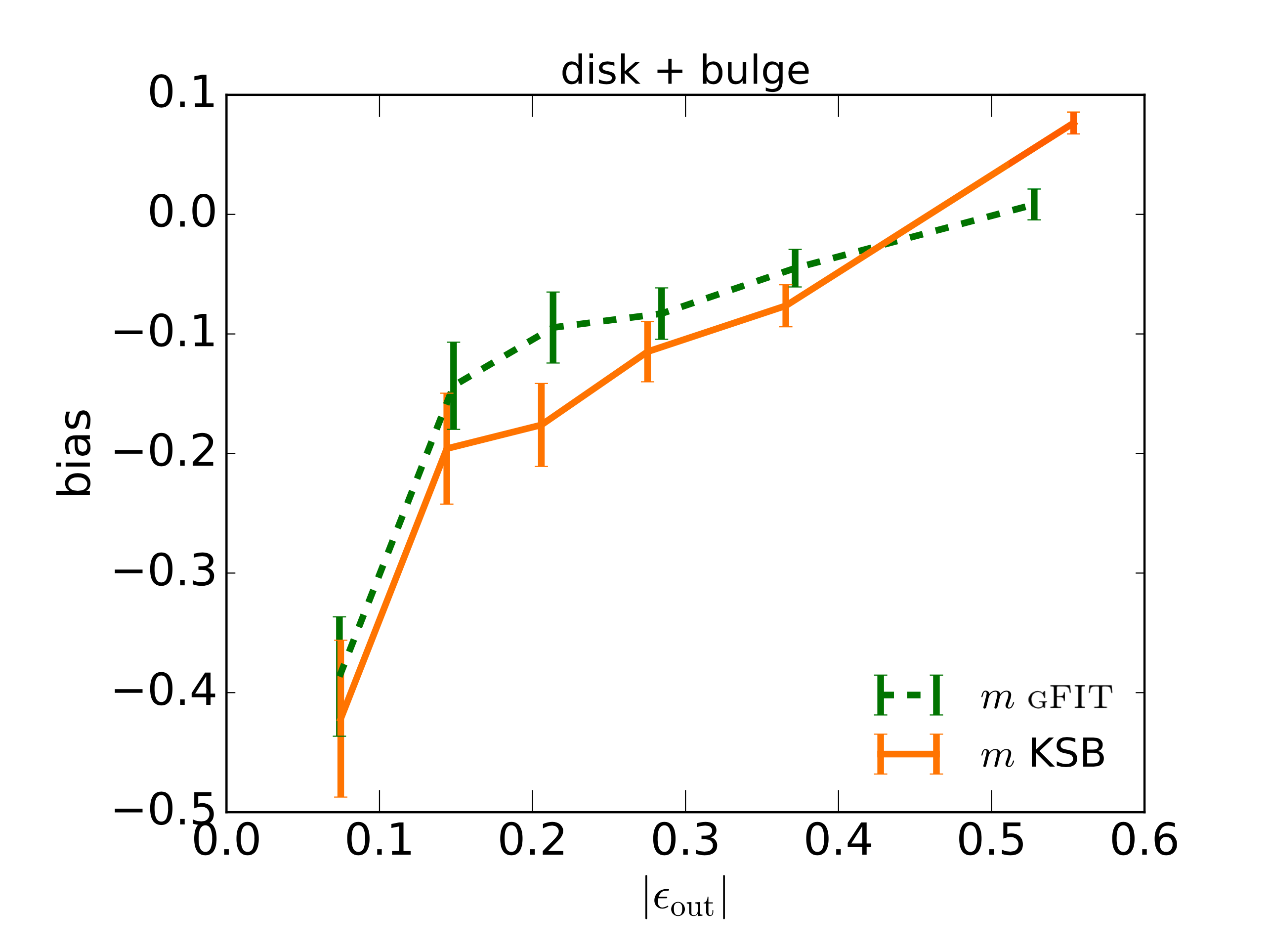}
\includegraphics[width=.45\linewidth]{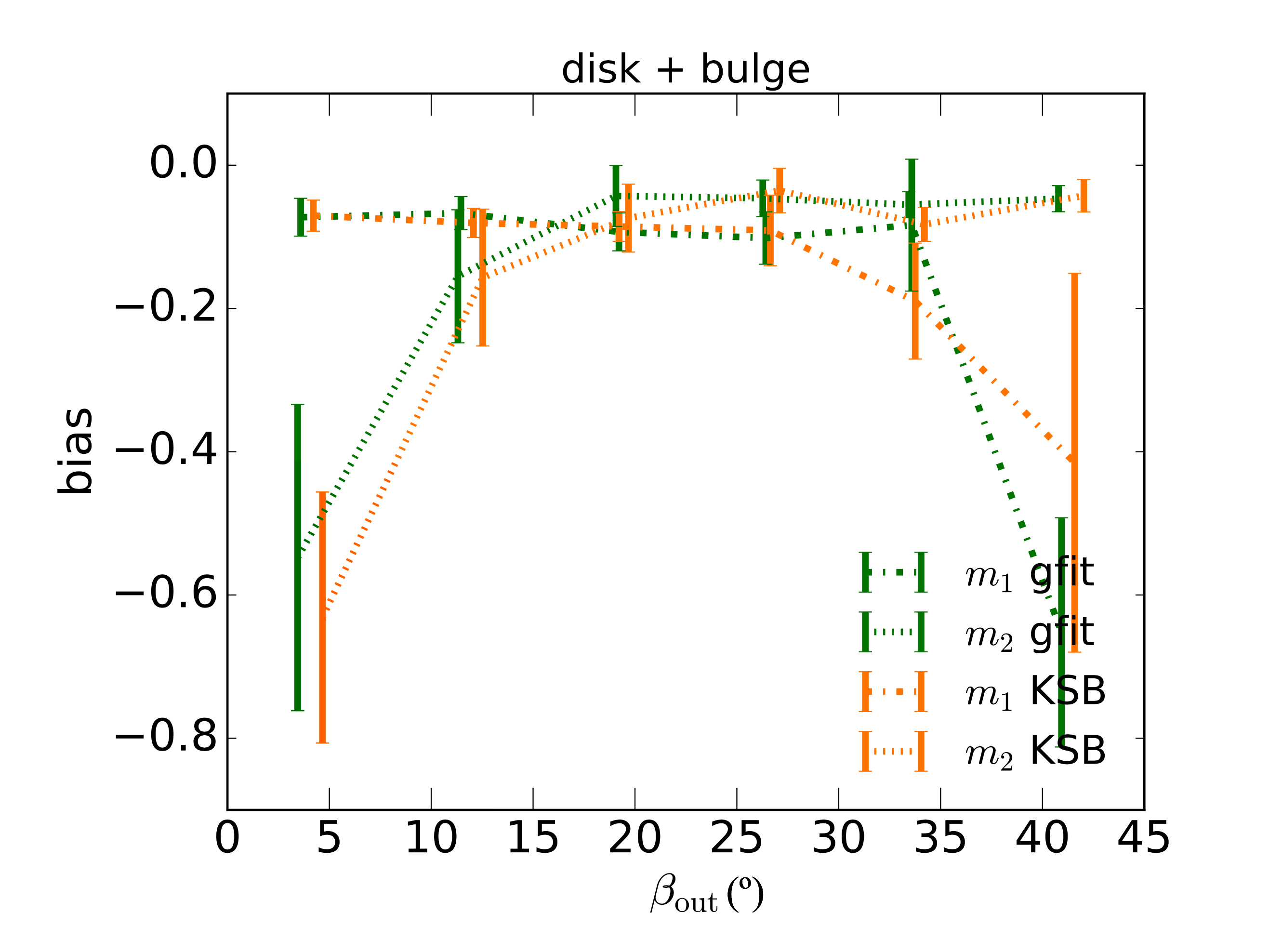}
\includegraphics[width=.45\linewidth]{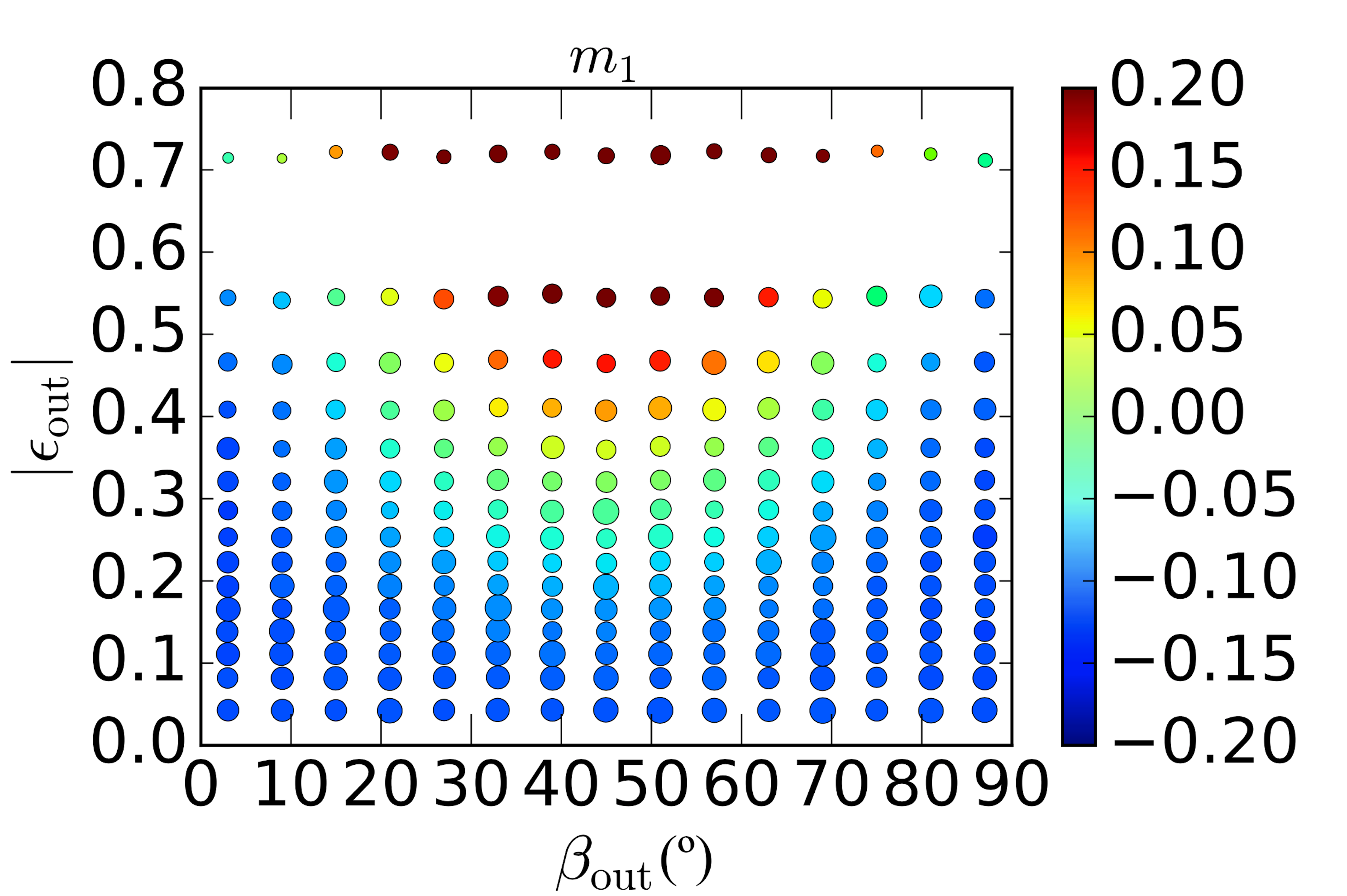}
\includegraphics[width=.45\linewidth]{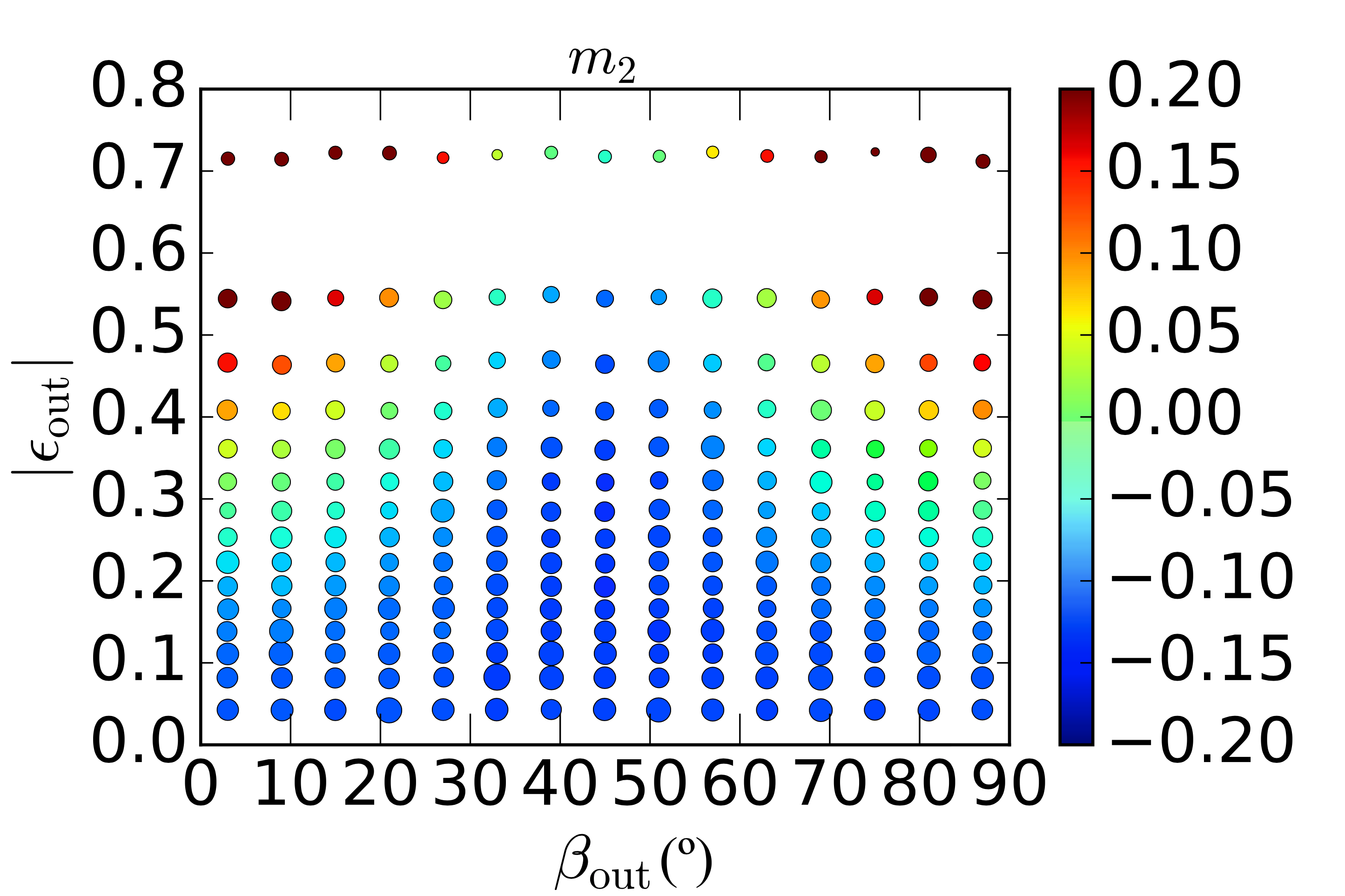}
\caption{Multiplicative shear bias as a function of the modulus of the observed ellipticity (top left panel) and the observed orientation angle (top right panel) for galaxies with a disc and a bulge. Green lines show the results of $m_1$ (dash-dotted), $m_2$ (dotted), and the mean $m$ (dashed) for \cgfit,\ and orange lines show $m_1$ (dash-dotted), $m_2$ (dotted), and the mean $m$ (solid) for \ksb.  The bottom panels show $m_1$ (left) and $m_2$ (right) represented by the colour code as a function of observed ellipticity and orientation for galaxies with a single bulge for \ksb. The point sizes are inversely proportional to the error on $m_{1,2}$ so that large points are more significant.}
\label{fig:out_plots}
\end{figure*}

In the previous sections, we explored the bias dependencies on input properties. The advantage of using the input properties to study shear bias is that we know precisely the relation between the images and these properties, but it has the handicap that they cannot be observed. On the other hand, measured properties are the information we can obtain from observations, which  can then be used for calibration. 

In \Fref{fig:out_plots}, we show shear bias as a function of the measured ellipticity (top left) and the orientation angle (top right) for galaxies with a bulge and a disc. We see very similar dependencies for both shear estimators. In the bottom panels, we show the dependence of $m_1$ (left) and $m_2$ (right) as a function of both the measured ellipticity and the measured orientation angle for the galaxies with only a bulge for \ksb.

We see that $m$ depends strongly on the measured modulus of the ellipticity,  $|\epsilon_{\rm out}|$, showing large biases for measured round objects that can be explained by the difficulties of defining the ellipticity of round objects. However,  small and dim galaxies are also difficult to measured correctly since they are more affected by noise and pixelization, being frequently but wrongly estimated as round. We also see a strong bias for the components that have been measured to be very small, so we see strong bias $m_1$ for galaxies with $\beta \sim 45^o$ and a strong bias $m_2$ for galaxies with $\beta \sim 0^o$. 
Finally, in the bottom panels we see that shear bias depends on both properties at the same time, and we cannot determine the shear bias of the galaxies if we only take into account one of the properties.

\begin{figure}
\centering
\includegraphics[width=.95\linewidth]{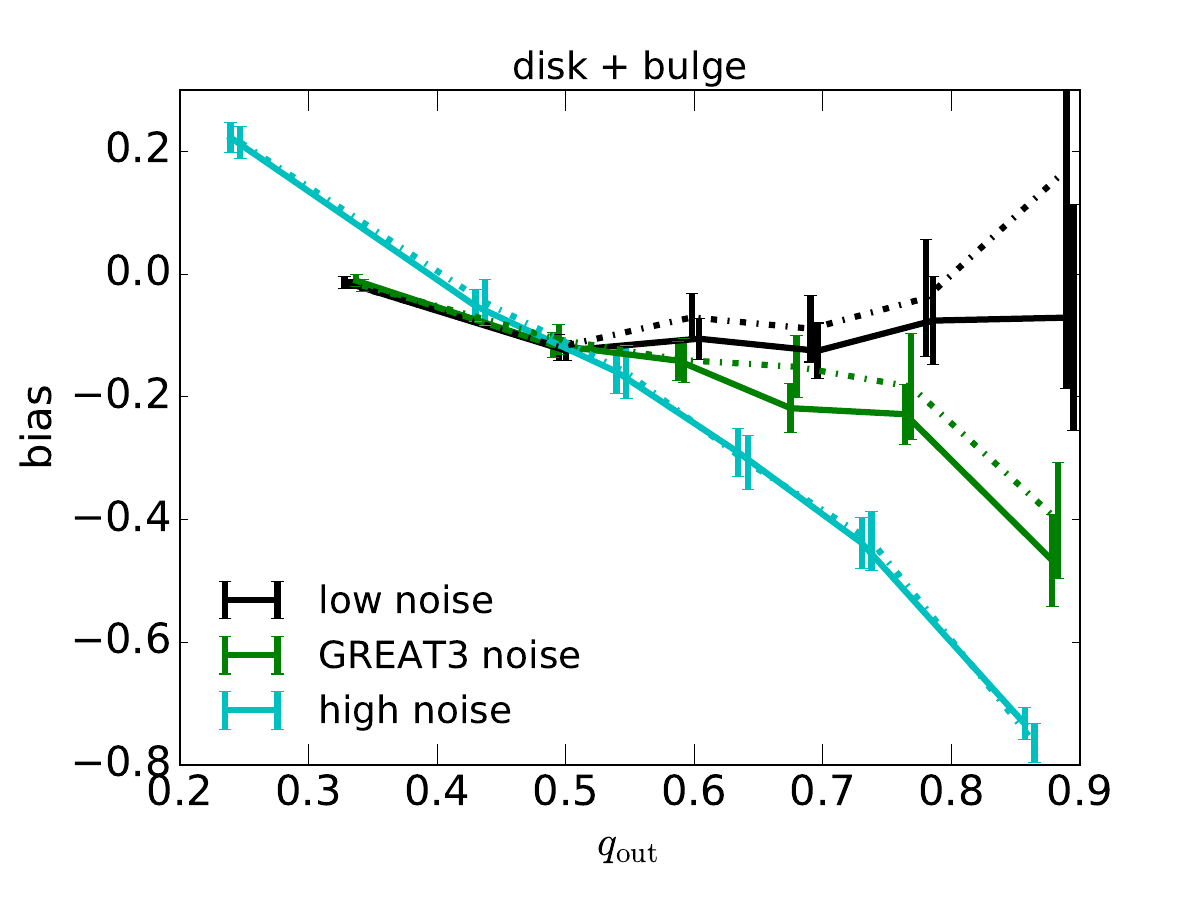}
\includegraphics[width=.95\linewidth]{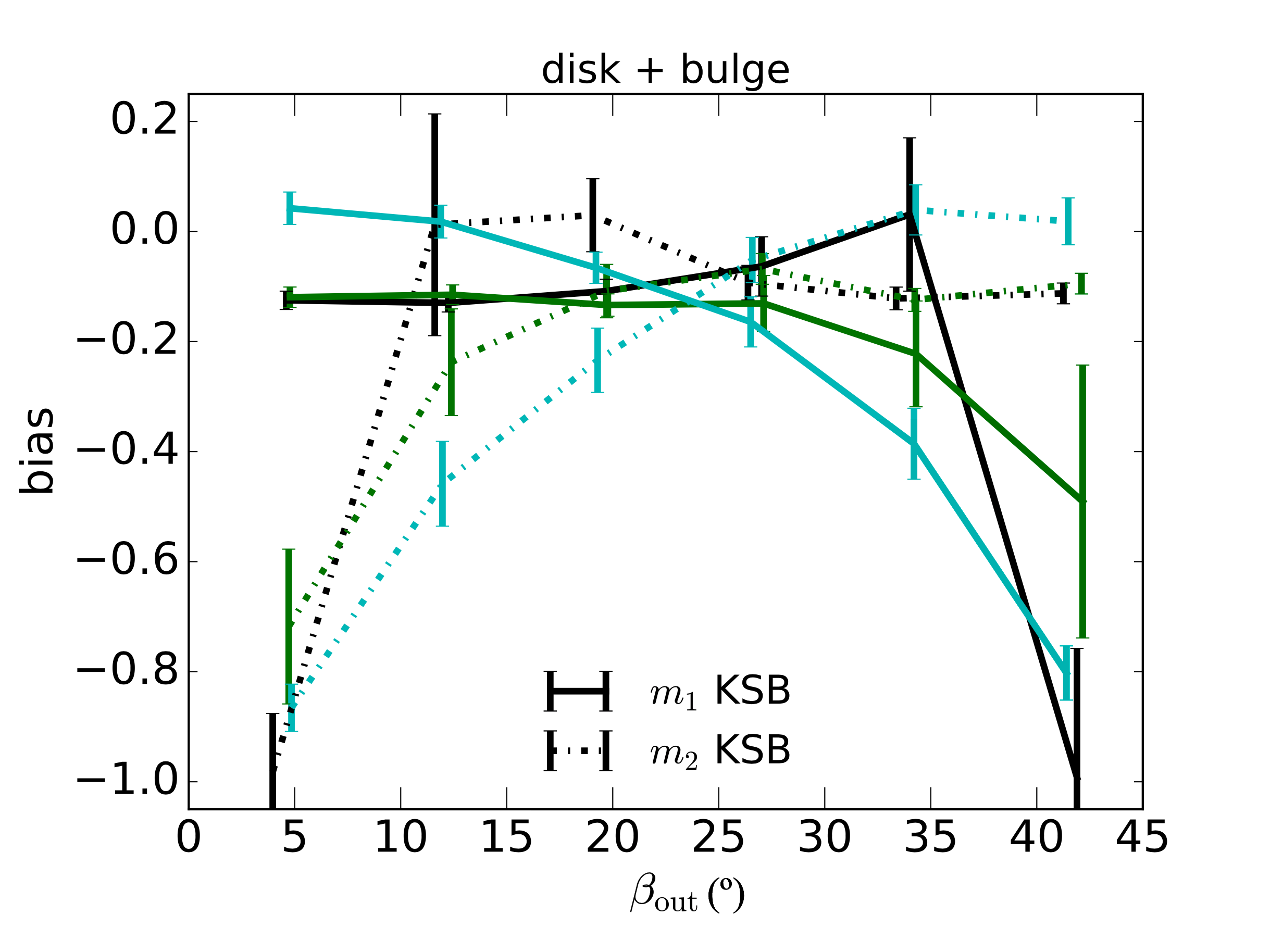}
\caption{Multiplicative shear bias as a function of the observed axis ratio (top) and orientation angle (bottom) using three different realizations with different noise variances. The green lines show the results for the original case, while in the cyan (black) lines show the results for the cases where we increased (decreased) the noise variance by a factor of 4. The results are only shown for \ksb,\ but we obtained very similar results for \gfit.}
\label{fig:out_q_noise}
\end{figure}

We find that noise is the main cause of these dependencies. To test this, we measured these bias dependencies by repeating the same analysis with the same images but generated with different levels of noise (changing the noise level by factors of $1/4$ and $4$). Here we only describe the results for \ksb,\ but the results are equivalent for \gfit.
In Fig. \ref{fig:out_q_noise}, we illustrate the impact of noise on the measured parameters $q$ (the axis ratio) and $\beta$.
We clearly see that noise has a strong impact on the shear bias of measured round objects. This is because, especially for small galaxies, noise contributes to the image with no correlated directions, so that galaxies will tend to be measured as rounder if the noise is strong enough. The bias for different orientations comes from the fact that small galaxies are more strongly affected by noise and pixelization. These results agree with \cite{Refregier2012} on the fact that noise bias affects galaxies with different ellipticity and morphologies differently.


\subsection{Differences between the shape estimators}

In the middle and bottom panels of \Fref{fig:in_plots_model}, we show the only significant difference found between \ksb\ and \cgfit\ given our set of simulations and the precision of the analysis. This shows shear bias as a function of the modulus of the intrinsic ellipticity. While \cgfit\ does not show a strong bias dependence on this property (especially for galaxies without a disc), \ksb\ shows a strong effect. This dependence from \ksb\ comes from the isotropic window function used in the method, as discussed in \cite{Viola2011}. However, in their study they show that different implementations of \ksb\ can produce different amplitudes of the bias as a function of the intrinsic ellipticity, and they propose different approximations to correct for it. In our study, we applied the implementation that showed the smallest bias dependence on ellipticity from those available in the public \shapelens\ repository.

It is not surprising that the different radial weightings of the methods give different shear bias results, as also stated in \cite{Bernstein2002} and \cite{Mandelbaum2017}. On the other hand, we have seen throughout our study that, apart from this case, most of the results are consistent between both estimators. This means that (almost) all the sources of bias found in this study do not come from the algorithm to estimate the shear, but from the image characteristics. 

\subsection{Other tests and results}\label{sec:other_tests}

In the previous sections, we show the most important bias dependences found from this study. Here we discuss some other tests where we do not see a strong impact on ellipticity or shear bias but are nevertheless worth mentioning.

First, we find additive bias to be weakly dependent on most of the properties, always being significantly smaller than multiplicative bias. Because of this, we only focus on multiplicative bias  in this work.

Second, we repeated the study from the simulated images; this time, we forced them to be centred in the stamps.  In this case, we ran \cgfit\ but kept the parameters of the centre positions fixed to the correct ones in the fitting process. We compared these results with the original case in order to see the impact of miscentring in this method and these images.
We do not find significant differences in the multiplicative bias, and for this reason we do not show the results in the paper. We find a small improvement on the amplitude of the additive bias, which makes it consistent with \ksb, indicating that \ksb\ is not affected by the miscentring like \cgfit\ is.
We also studied the $+$ and $\times$ components of shear and ellipticity bias, but we find consistent results with respect to the $1$ and $2$ components (except, as expected, for the properties related to orientation, where the $+$ and $\times$ components are uncorrelated with the pixel directions).

Finally, we studied the effect of the PSF by repeating the same test from the same images but applying an isotropic Gaussian PSF. The bias dependencies found are the same in both cases, and the PSF anisotropies only affect the error bars of our measurements. This result can be expected from the fact that we used the knowledge of the PSF for the shape estimation in both methods; therefore, this test does not reflect the errors of the methods originating from untracked PSF errors and their anisotropy. 

\subsection{Comparison with existing literature}

Some of the results from this paper confirm previous analyses. In addition, we find new dependencies of shear bias that, to our knowledge, have not been studied before or to which existing work cannot be directly compared. In this section, we specify the main similarities and differences with respect to existing literature as well as the new results shown in this paper.

Many studies have analysed the average shear bias over an ensemble of galaxies with different properties with a focus on comparing different shear estimators, such as the STEP and GREAT challenges \citep{Heymans2006,Bridle2009,Kitching2012,Mandelbaum2015}. In those analyses, different estimators showed different results, up to a $5-10\%$ multiplicative bias for most of the methods. Although the comparison with our case is non-trivial due to the differences in the estimators and the image properties used, our results seem consistent with those previous results. In particular, both estimators, \gfit\ and \ksb,\ have an average bias of $\sim5\%$, which was expected from the results from the GREAT3 CSC branch \citep[see Fig. 17 and table D2 of][]{Mandelbaum2015}. The negative sign and magnitude of the \ksb\ bias is in agreement with \cite{Hoekstra2017}, who show a negative multiplicative bias of $2-6\%$ throughout their analysis.  The $5\%$ bias for the \ksb\ method is also consistent with the \ksb\ performance in the GREAT10 challenge \citep{Kitching2012}. The \gfit\ method shows a different performance in that study compared to ours, but the comparison is misleading since in GREAT10 \gfit\ was sensitive to a truncation effect that we avoided by using larger postage stamps.
We find that shear bias decreases with the size and flux of the galaxies, going from a $10\%$ bias for dim and small objects to almost no bias for the largest (brightest) ones. These results are consistent with previous studies \citep{Massey2007b,Hoekstra2015}, although the amplitudes of the effects can vary because of the differences in the estimators and the characteristics of the image simulations.

Most of the previous studies either focus on one estimator and explore the shear bias dependencies on input properties, or compare different estimators of the average shear bias over all galaxy images. Here, we compare two common but different estimators analysing, at the same time, the shear bias dependences on both input and measured parameters. One of the interesting results from this paper is the fact that both \gfit\ and \ksb\ give (with a few expected exceptions) very consistent results on most of the property dependencies. This is not necessarily expected given the different nature of both estimators (MLE versus moment-based).
We have shown the bias dependence on the S\'ersic index $n$, showing different trends than in \cite{Kacprzak2014}. In that paper, the authors used an MLE but assumed a different model (single S\'ersic) than \gfit. These comparisons show how sensitive the assumed model is to the shear bias dependence on $n$.
We show, for the first time, the orientation angle dependence of shear bias, giving antisymmetric dependencies for $m_1$ and $m_2$. This has not been explored in previous analyses, although \cite{Kacprzak2012} show an indication of ellipticity bias that changes with the orientation angle in their Fig. A1. We checked that we obtained a similar behaviour for the ellipticity bias. Finally, most of the studies show the bias dependencies as a function of input parameters, while they can never be obtained for real images. Moreover, measured and input properties can show very different biases. In this paper, we show the different behaviour between input and measured morphology. In Figs. \ref{fig:out_plots} and \ref{fig:out_q_noise}, we show the shear bias dependencies on measured ellipticity and orientation angle at the same time, together with the impact of noise.

\section{Conclusions}\label{sec:conclusions}

In this study we explored the dependencies of ellipticity and shear bias on input properties of simulated galaxy images, output properties, and the impact of noise, pixelization, and PSF anisotropy for two different shape estimators. We used \galsim\ to simulate the images of galaxies from the GREAT3 CSC \citep{Mandelbaum2014} parameters, and we compared the ellipticity and shear bias obtained from the MLE \gfit\ \citep{Gentile2012,Mandelbaum2015} and the moment-based \ksb\ method, which is available from the public software \shapelens\ \citep{Viola2011}. In order to study the effects of pixelization, noise, and PSF anisotropy, we repeated the analysis with some variations from the original simulated images, where we tested different levels of noise variance and an isotropic Gaussian PSF. In this paper, we focused on multiplicative bias since, given the precision of the analysis, we do not find important dependencies on additive bias. Here we discuss the most important conclusions from our study.

First, we find a good agreement between the \gfit\ and \ksb \ shape estimators. Given the differences in the nature of these two estimators, this suggests that most of the dependencies found in this paper can be common for all image characteristics and for many shape estimators based on moments or model fitting.

Second, we show that ellipticity bias and shear bias present very different behaviours since they reflect the different sensitivities of the shape estimators. 
These differences agree with previous studies and show the importance of studying the average shear bias and not the per-galaxy ellipticity bias for cosmological analysis.


Finally, we studied the dependencies of bias on all input and output properties, and we determined the ones to which bias is most sensitive. We find three types of dependencies:
\begin{itemize}
\item Size and shape dependency: Shear and ellipticity bias depends on the properties related to the dimensions of the galaxy image, such as the bulge and disc fluxes and sizes. Bias is larger for small objects since their shape is more difficult to measure. Round galaxies also show large ellipticity biases because the measurement of the ellipticity is strongly affected by pixelization and noise. Elliptical and large images are less sensitive to these aspects and thus show a smaller bias.
\item Orientation dependence: Shear bias depends strongly on orientation, with asymmetric dependencies for $m_1$ and $m_2$. This is due to pixelization effects that make the estimation of the ellipticity more sensitive to small rotations than to small elongations along the pixel directions.
\item Model bias: Shear bias is larger for galaxies containing a bulge and a disc than for galaxies consisting of only a bulge. We also find a strong dependence of shear bias on the S\'ersic index $n$. 
Even though the \ksb\ and \cgfit\ use different assumptions or treatments regarding the luminosity profile of the images, the model bias for these galaxies is similar.
\end{itemize}
We find that the bias as a function of measured ellipticity and orientation is strongly affected by noise. This is because galaxies that are strongly affected by noise can be systematically interpreted to have the same properties even if their input properties are different. 


The results and conclusions of this paper are limited to the accuracy that we can reach with the simulation images used. More optimal approaches to estimate shear bias \citep{Pujol2019} or using a larger set of images would help to improve the analysis and potentially find other smaller dependencies or differences between the estimators. However, this study has been useful to identify the main causes of shear bias and the properties on which bias is most dependent. We also highlight the complexity of these dependencies, the impact of the coupling between different properties on shear bias, and the need to study several properties simultaneously in order to have a better understanding of the nature of shear bias. This strongly motivates us to develop a machine learning tool to characterize shear bias that simultaneously takes many properties and dependencies  into account for a more precise calibration approach. This new method is presented in detail in Paper II \citep{Pujol2020}.

\section*{Acknowledgements}
AP, FS and JB acknowledge support from a European Research Council Starting Grant (LENA-678282).
FC is supported by the Swiss National Science Foundation.
MG acknowledges the support of the Swiss State Secretariat for Education, Research and Innovation SERI and ESA's PRODEX programme.

\bibliography{biblist}

\appendix

\end{document}